\documentclass[apj]{emulateapj}
\usepackage{graphics,graphicx,times,natbib,longtable,placeins,mathrsfs}

\newcommand{\sersic}{{S\'{e}rsic }}

\def\mstar{${\rm M_{*}}$}

\def\logm{{\rm log(M_{*}/M_\odot)}}

\def\fclumpy{$f_{clumpy}$}
\def\csfr{$C_{SFR}$}
\def\cuv{$C_{UV}$}
 
\def\hst{{\it HST}}

\def\spose#1{\hbox to 0pt{#1\hss}}
\def\lta{\mathrel{\spose{\lower 3pt\hbox{$\mathchar"218$}}
     \raise 2.0pt\hbox{$\mathchar"13C$}}}

\shorttitle{UV-bright Clumps at $0.5<z<3$}
\shortauthors{Guo et al.}

\begin{document}

\title{Clumpy Galaxies in CANDELS. I. The Definition of UV Clumps and the
Fraction of Clumpy Galaxies at $0.5<z<3$}

\author{Yicheng Guo$^{1}$, Henry C. Ferguson$^{2}$, Eric F. Bell$^{3}$, David
C. Koo$^{1}$, Christopher J. Conselice$^{4}$, Mauro Giavalisco$^{5}$, Susan
Kassin$^{2}$, Yu Lu$^{6}$, Ray Lucas$^{2}$, Nir Mandelker$^{7}$, Daniel M.
McIntosh$^{8}$, Joel R. Primack$^{9}$, Swara Ravindranath$^{2}$, Guillermo
Barro$^{1}$, Daniel Ceverino$^{10}$, Avishai Dekel$^{7}$, 
Sandra M. Faber$^{1}$, Jerome J. Fang${^1}$, Anton M. Koekemoer$^{2}$, 
Kai Noeske$^{2}$, Marc  Rafelski$^{11,12}$, Amber Straughn$^{12}$}
\affil{$^1$ UCO/Lick Observatory, Department of Astronomy and Astrophysics, University of California, Santa Cruz, CA, USA; {\it ycguo@ucolick.org}}
\affil{$^2$ Space Telescope Science Institute, Baltimore, MD, USA}
\affil{$^3$ Department of Astronomy, University of Michigan, Ann Arbor, MI, USA}
\affil{$^4$ School of Physics and Astronomy, University of Nottingham, University Park, Nottingham NG7 2RD, UK}
\affil{$^5$ Department of Astronomy, University of Massachusetts, Amherst, MA, USA}
\affil{$^6$ Kavli Institute for Particle Astrophysics and Cosmology, Stanford, CA, USA}
\affil{$^7$ Center for Astrophysics and Planetary Science, Racah Institute of
Physics, The Hebrew University, Jerusalem 91904 Israel}
\affil{$^8$ Department of Physics and Astronomy, University of Missouri-Kansas City, Kansas City, MO, USA}
\affil{$^9$ Department of Physics, University of California, Santa Cruz, CA, USA}
\affil{$^{10}$ Departamento de F\'isica Te\'orica, Universidad Aut\'onoma de Madrid, 28049 Madrid, Spain}
\affil{$^{11}$ NASA Postdoctoral Program Fellow}
\affil{$^{12}$ Astrophysics Science Division, Goddard Space Flight Center, Code 665, Greenbelt, MD 20771, USA}



\begin{abstract} Although giant clumps of stars are thought to be crucial to
galaxy formation and evolution, the most basic demographics of clumps 
are still uncertain, mainly because the definition of clumps has not been
thoroughly discussed. In this paper, we carry out a study of the basic
demographics of clumps in star-forming galaxies at $0.5<z<3$, using our
proposed physical definition that UV-bright clumps are discrete
star-forming regions that individually contribute more than 8\% of the
rest-frame UV light of their galaxies. 
Clumps defined this way are significantly brighter than the HII regions of 
nearby large spiral galaxies,
either individually or blended, when physical spatial resolution and
cosmological dimming are considered. Under this definition, we measure the
fraction of star-forming galaxies that have at least one off-center clump
($f_{clumpy}$) and the contributions of clumps to the rest-frame UV light and
star formation rate (SFR) of star-forming galaxies in the CANDELS/GOODS-S and
UDS fields, where our mass-complete sample consists of 3239 galaxies with axial
ratio $q>0.5$. The redshift evolution of $f_{clumpy}$ changes with the stellar
mass (\mstar) of the galaxies. Low-mass ($\logm<9.8$) galaxies keep an almost
constant $f_{clumpy}$ of $\sim$60\% from $z\sim3$ to $z\sim0.5$.
Intermediate-mass and massive galaxies drop their $f_{clumpy}$ from 55\% at
$z\sim3$ to 40\% and 15\%, respectively, at $z\sim0.5$.
We find that (1) the trend of disk stabilization predicted by violent disk
instability matches the \fclumpy\ trend of massive galaxies;
(2) minor mergers are a viable explanation of the
\fclumpy\ trend of intermediate-mass galaxies at $z<1.5$, given a realistic
observability timescale; and (3) major mergers are unlikely responsible for the
\fclumpy\ trend in all masses at $z<1.5$. 
The clump contribution to the rest-frame UV light of star-forming galaxies
shows a broad peak around galaxies with $\logm \sim10.5$ at all redshifts. 
The clump contribution in the intermediate-mass and massive galaxies is
possibly linked to the molecular gas fraction of the galaxies. The clump
contribution to the SFR of star-forming galaxies, generally around 4--10\%,
also shows dependence on the galaxy \mstar, but for a given galaxy \mstar, its
dependence on the redshift is mild.  \end{abstract}

\section{Introduction}
\label{intro}

The emergence of facilities with high sensitivity and high resolution, e.g.,
\hst/ACS, NICMOS, and WFC3, enables astronomers to resolve galaxy morphology
and structure to kpc scale to study the properties of galactic
sub-structures at high redshift
\citep[e.g.,][]{elmegreen05,elmegreen07,elmegreen09a,elmegreen09b,gargiulo11,szomoru11,ycguo11peg,ycguo12clump}.
An important observational feature of high-redshift star-forming galaxies
(SFGs) is the existence of giant kpc-scale clumps of stars or star formation
activities
\citep[e.g.,][]{conselice04,elmegreen05,elmegreen07,elmegreen09a,bournaud08,genzel08,genzel11,fs11b,ycguo12clump,wuyts12},
which are unusual in massive low-redshift galaxies.

The giant clumps are mostly identified in the deep and high-resolution
rest-frame UV images \citep[e.g.,][]{elmegreen05,elmegreen07,ycguo12clump} and
rest-frame optical images \citep[e.g.,][]{elmegreen09a,fs11b}. They are also
seen in the rest-frame optical line emission from NIR integral field
spectroscopy \citep[e.g.,][]{genzel08,genzel11} or CO line
emission of lensed galaxies \citep[e.g.,][]{jones10,swinbank10}. The typical
stellar mass (\mstar) of clumps is ${\rm 10^{7}-10^{9} M_\odot}$
\citep[e.g.,][]{elmegreen07,ycguo12clump}, and the typical size is $\sim$1 kpc
or less \citep[e.g.,][]{elmegreen07,fs11b,livermore12}. The clumps have blue
UV--optical colors and are shown to be regions with enhanced specific star
formation rates (SSFR), which are higher than that of their surrounding areas
by a factor of several \citep[e.g.,][]{ycguo12clump,wuyts12,wuyts13}.  Both
morphological analysis \citep[e.g., the \sersic models,][]{elmegreen07} and gas
kinematic analysis \citep[e.g., H$\alpha$ velocity maps,][]{genzel08,genzel11}
show that many clumpy galaxies have underlying disks.

%

Although clumps are thought to be important laboratories to test our knowledge
of star formation, feedback, and galactic structure formation, the definition
of ``clump'' has not been thoroughly discussed. Clumps were originally defined
through the appearance of galaxies by visual inspection
\citep[e.g.,][]{cowie95,vandenbergh96,elmegreen04,elmegreen05,elmegreen07}.
Visual definitions are, however, subjective and hard to reproduce. More and
more studies have begun to automate the clump detection
\citep[e.g.,][]{conselice03,conselice04,fs11b,ycguo12clump,wuyts12,murata14}.
Although these automated detections are easier to reproduce and to apply to
large samples, most of them define clumps based on the appearance of galaxies,
namely, the intensity contrast between the peak and the local background in
galaxy images. The biggest problem of such definitions is that the appearance
of even the same type of galaxies changes with the sensitivity and resolution
of observations.  Therefore, each of such definitions of clumps is actually
bound to a given observation, which makes comparisons between different
observations difficult. As a result, there are still large uncertainties in the
most basic demographics of clumps: what fraction of SFGs have clumps, and what
fraction of the total star formation occurs in clumps. 

The measurement of the fraction of clumpy galaxies in the overall sample of
SFGs (\fclumpy\ ) shows a large dispersion in literature.
\citet{ravindranath06} claimed that clumpy galaxies are about 30\% of the
population at z$\sim$3, while \citet{elmegreen07} argued that the dominant
morphology for z$\gtrsim$2 starbursts is clumpy galaxies. \citet{ycguo12clump}
found a high \fclumpy\ $\sim$ 67\% for SFGs at z$\sim$2 in HUDF. However, their
sample contains only 15 galaxies, which may be biased toward bright, blue, and
large galaxies because they include only spectroscopically observed galaxies.
\citet{wuyts12} measured the fraction of clumpy galaxies in a mass-complete
sample of SFGs at z$\sim$2 by using multi-waveband images and \mstar\ maps.
They found that the clumpy fraction depends sensitively on the light/mass map
used to identify the regions with excess surface brightness and decreases from
about 75\% for galaxies selected through rest-frame 2800\AA\ images to about
40\% for those selected through rest-frame V-band images or \mstar\ maps.  

The clump contribution to the UV light and star formation rate (SFR) of the
galaxies is closely related to the physics that drives galaxy formation and
evolution, e.g., gas accretion rate, gas fraction, and star formation
efficiency. For example, if clumps are formed in-situ through the violent disk
instability \citep[VDI;][]{dekel09} in gas-rich rotating disks that are
perturbed by the accreted gas inflow, the clump contribution is expected to
drop from high redshift to low redshift, because the cosmic cold gas accretion
quickly declines with the cosmic time \citep[e.g.,][]{keres05,dekel09gas}.
\citet{wuyts12, wuyts13}, using color/SFR excess to identify clumpy regions,
found the clump contribution to the cosmic SFR decreasing from $z=2.5$ to
$z=1$, consistent with the above prediction. On the other hand,
\citet{mandelker14}, using gas maps to identify clumps, found that the clump
contribution to SFR in the \citet{ceverino10} numerical simulations is almost
flat, if not increasing, from $z=3$ to $z=1$. A possible reason for the
discrepancy is that the above two studies did not define clumps in the same
way. In fact, \citet{moody14} found that the one-to-one correspondence among
the clumps defined through gas, young stars, and mass is poor. To unify the
current rapid emergence of multi-wavelength observations of clumps as well as
the state-of-the-art numerical simulations, it is crucial to have a more
physical definition of ``clump'' to move the studies of clumps and clumpy
galaxies forward.

In this paper, we propose a definition of clumps based on their intrinsic
rest-frame UV properties and present a comprehensive measurement of \fclumpy\
and its variation with redshift and \mstar, exploiting the advantage of high
resolution and deep sensitivity of \hst/ACS and WFC3 images in the
CANDELS/GOODS-S and UDS fields. We also measure the clump contribution to the
rest-frame UV light and SFR of SFGs.

The paper is organized as follows. The data and sample selection are presented
in Sec. \ref{data}. In Sec. \ref{clfinder}, we start our clump definition by
the traditional way of detecting discrete star-forming regions through the
intensity contrast between the peak and background of galaxy images. We use an
automated algorithm to detect the star-forming regions in the same rest-frame
UV bands across a wide redshift range of $0.5<z<3.0$.  In Sec. \ref{fuv}, we
measure the incompleteness-corrected fractional luminosity function (FLF),
namely, the number of star-forming regions per galaxy that contribute a given
fraction of the total UV light of the galaxies.  In Sec. \ref{definition}, we
compare the FLF of redshifted nearby galaxies with that of real galaxies.  This
comparison allows us to define ``clumps'' as star-forming regions whose
fractional luminosity (FL) is significantly higher than that of redshifted
nearby star-forming regions.  Given this definition, we measure \fclumpy\ in
Sec. \ref{fclumpy} and the clump contribution to the UV light and SFR of SFGs
in Sec. \ref{clumpfuv}. Conclusions and discussions will be presented in Sec.
\ref{conclusion}.

Throughout the paper, we adopt a flat ${\rm \Lambda CDM}$ cosmology with
$\Omega_m=0.3$, $\Omega_{\Lambda}=0.7$ and use the Hubble constant in terms of
$h\equiv H_0/100 {\rm km~s^{-1}~Mpc^{-1}} = 0.70$.  All magnitudes in the paper
are in AB scale \citep{oke74} unless otherwise noted.

\section{Data and Sample Selection} \label{data}

\subsection{Catalogs and Images} \label{data:cat}

The sample of galaxies used in this paper is selected from the CANDELS/GOODS-S
and UDS fields \citep{candelsoverview,candelshst}. CANDELS (HST-GO-12060) has
observed both fields with the \hst/WFC3 F160W band, reaching a 5$\sigma$
limiting depth (within a 0\farcs17-radius aperture) of 27.36, 28.16, and 27.35
AB mag for the GOODS-S wide ($\sim$1/3 of the GOODS-S), deep  ($\sim$1/3 of the
GOODS-S), and UDS fields. The remaining 1/3 of the GOODS-S field has the F160W
observation from ERS with a depth similar to that of the GOODS-S/deep region.
Based on the source detection in the F160W band, the CANDELS team has made a
multi-wavelength catalog for each field, combining the newly obtained CANDELS
HST/WFC3 data with existing public ground-based and space-based data. The
details of the catalogs are given by \citet[][for GOODS-S]{ycguo13goodss} and
\citet[][for UDS]{galametz13uds}. In brief, \hst\  photometry was measured by
running SExtractor on the point spread function (PSF)-matched images in the dual-image mode, with the
F160W image as the detection image. Photometry in ground-based and IRAC images,
whose resolutions are much lower than that of the F160W images, was measured by
using TFIT \citep{laidler07}, which fit the PSF-smoothed high-resolution image
templates to the low-resolution images to measure the fluxes in the
low-resolution images.

Clumps are detected from the \hst/ACS images of the galaxies. The spatial
resolution of the ACS images (0\farcs1--0\farcs12) is equivalent to $\sim$1 kpc
in our target redshift range. In GOODS-S, the images are the latest mosaics of
the \hst/ACS F435W, F606W, and F775W bands from the GOODS Treasury Program.
They consist of data acquired prior to the \hst\ Servicing Mission 4, including
mainly data from the original GOODS \hst/ACS program in \hst\ Cycle 11
\citep[GO 9425 and 9583; see][]{giavalisco04goods} and additional data acquired
on the GOODS fields during the search for high redshift Type Ia supernovae
carried out during Cycles 12 and 13 \citep[Program ID 9727, P.I. Saul
Perlmutter, and 9728, 10339, 10340, P.I. Adam Riess; see, e.g.,][]{riess07}.
The 5$\sigma$ limiting depths (within a 0\farcs17-radius aperture) of ACS
F435W, F606W, and F775W bands in the GOODS-S field are 28.95, 29.35, and 28.55
AB, respectively. In UDS, CANDELS has taken parallel observations on the F606W
and F814W bands, with the 5$\sigma$ limiting depths of 28.49 and 28.53 AB.

Besides doubling our sample size, using both the GOODS-S and UDS fields allows
us to evaluate the incompleteness of clump detections at different observation
depths. 

\subsection{Galaxy Properties}
\label{data:prp}

The properties of galaxies in the two fields are measured through fitting the
broad-band spectral energy distributions (SED) in the catalogs to synthetic
stellar population models. We use the official CANDELS photometric redshift
(photo-z) catalogs in the two fields, which combine the results from more than
a dozen photo-z measurements with various SED-fitting codes and templates. The
technique is fully described in \citet{dahlen13}. 
Stellar mass and other stellar population properties (such as age, extinction,
UV-based SFR, etc.) are measured by using FAST \citep{kriek09fast}, with
redshift fixed to the best available ones (spectroscopic or photometric). The
modeling is based on a grid of \citet{bc03} models that assume a
\citet{chabrier03} IMF, solar metallicity, exponentially declining star
formation histories, and a Calzetti extinction law
\citep{calzetti94,calzetti00}. 

SFRs are measured on a galaxy-by-galaxy basis using a ladder of SFR indicators
as described in \citet{wuyts11a}. The method essentially relies on IR-based SFR
estimates for galaxies detected at mid- to far-IR wavelengths, and SED-modeled
SFRs for the rest. As shown in \citet{wuyts11a} the agreement between the two
estimates for galaxies with a moderate extinction (faint IR fluxes) ensures the
continuity between the different SFR estimates. For IR-detected galaxies the
total SFRs, SFR IR+UV, were then computed from a combination of IR and
rest-frame UV luminosity (uncorrected for extinction) following
\citet{kennicutt98}. We refer readers to \citet{barro11b, barro14a} for the
details of our measurements of galaxy properties.

\begin{figure}[htbp] \center{\hspace{-0.2cm}
\includegraphics[scale=0.9, angle=0]{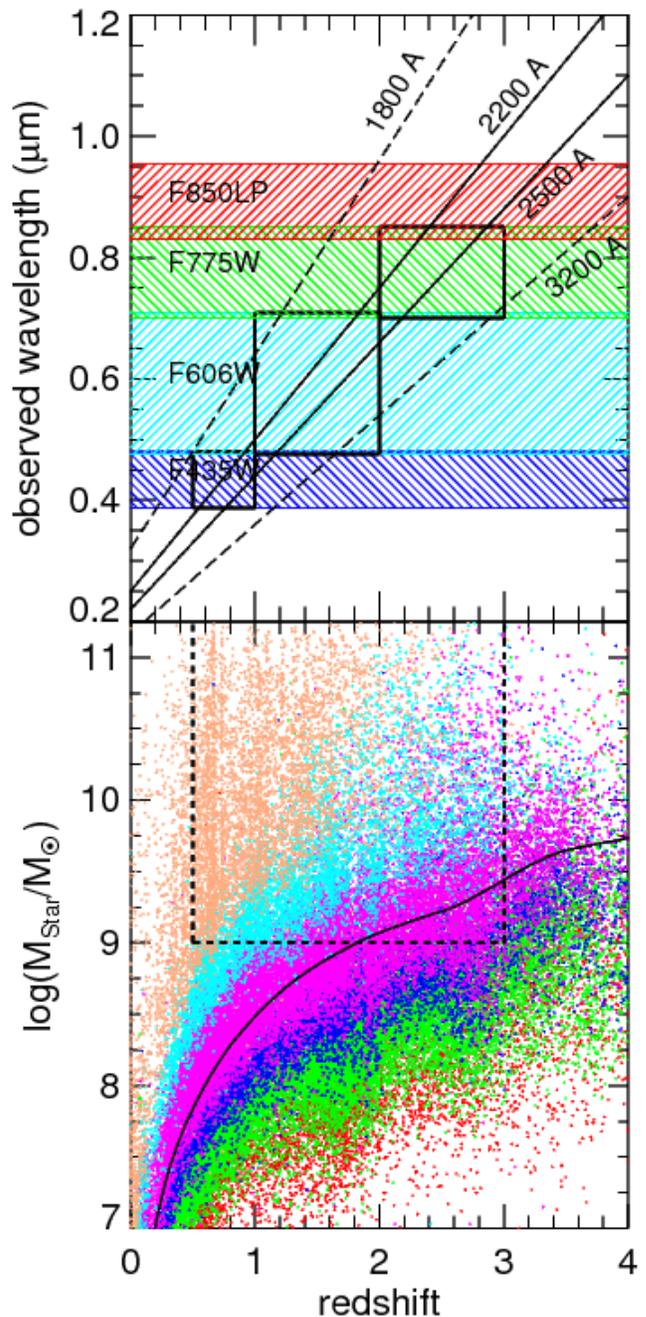}} 
\caption[]{{\it Upper:} Shift of our clump detection bands. The colored areas
show the wavelength coverage of ACS filters. The large black rectangles show
the bandpass used for detecting clumps in each redshift range in our study.
Black lines, from bottom to top, show the observed wavelength of rest-frame
1800, 2200, 2500, and 3200 \AA, respectively.  {\it Lower:} \mstar--redshift
diagram of the CANDELS/GOODS-S catalog. Galaxies with $H_{F160W}\geq27.0$
(red), $26.0\leq H_{F160W}<27.0$ (green), $25.0\leq H_{F160W}<26.0$ (blue),
$24.0\leq H_{F160W}<25.0$ (purple), $23.0\leq H_{F160W}<24.0$ (cyan), and
$H_{F160W}\leq23.0$ (light brown) are shown. The black curve shows the \mstar
of an SED template with $H_{F160W}=24.5$ AB and a constant star formation
history over an age of 0.5 Gyr at different redshifts. Black dashed lines show
the boundary of our sample.
\label{fig:hmagcut}}
\vspace{-0.4cm}
\end{figure}

\begin{figure*}[htbp]
\center{\hspace*{-0.5cm}\includegraphics[scale=0.6, angle=0]{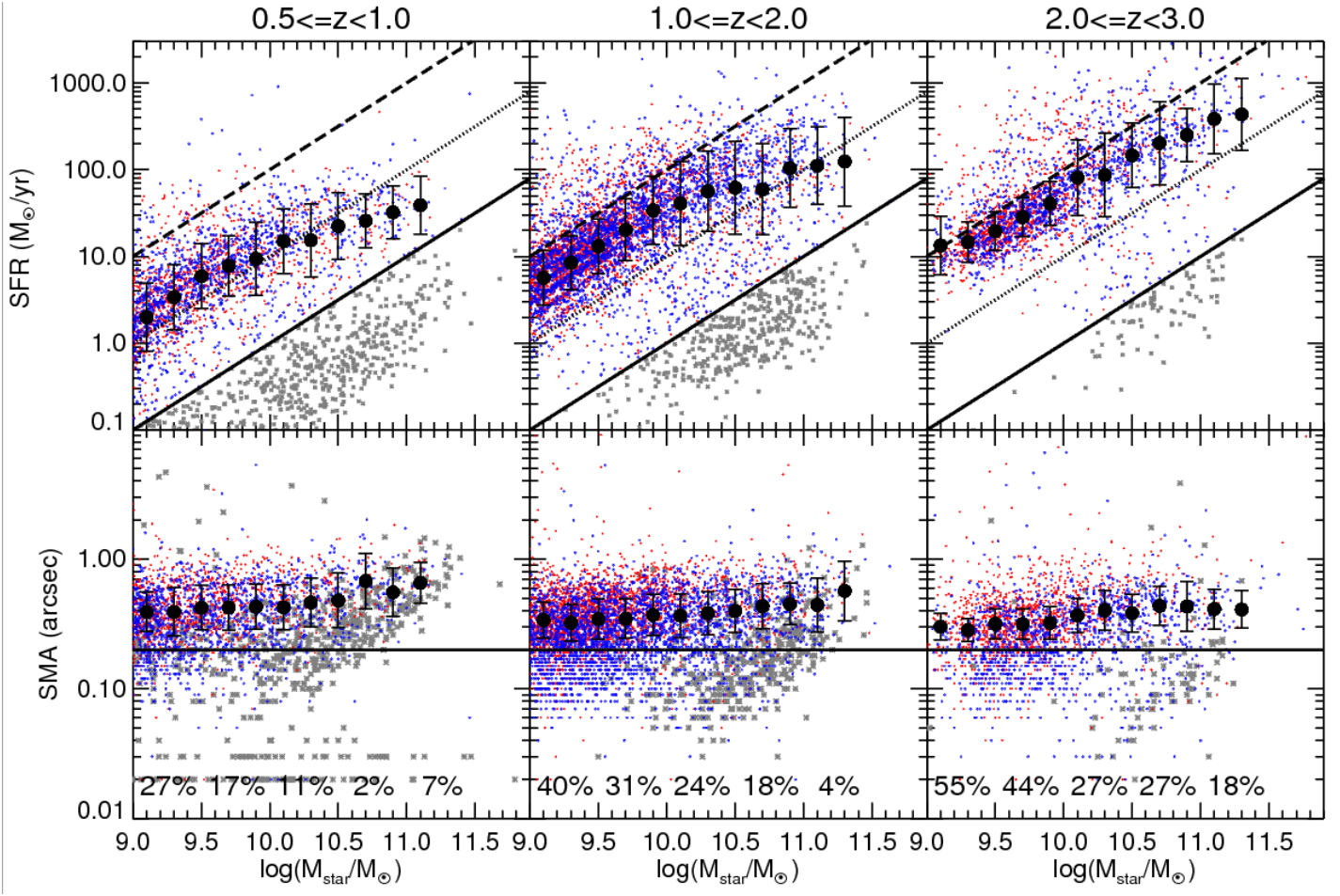}}
\caption[]{Sample selection. Galaxies in the CANDELS/GOODS-S and UDS with
$H_{F160W} < 24.5$ AB are plotted in the SFR--\mstar\ and semi-major axis
(SMA)--\mstar\ diagrams. Galaxies with SSFR$> 0.1 Gyr^{-1}$ and axial ratio
$q>0.5$ (blue) are selected into our sample of clump detection.  Red points
show galaxies with SSFR$> 0.1 Gyr^{-1}$ but $q\leq0.5$, while gray points show
galaxies with SSFR$\leq0.1 Gyr^{-1}$. Black filled circles with error bars show
the median and scatter of the star-forming galaxies (with SSFR$> 0.1 Gyr^{-1}$)
in the SFR--\mstar\ diagram. Black solid, dotted, and dashed lines in the upper
panels show the relations of SSFR=0.1, 1, and 10 $Gyr^{-1}$. Black horizontal
lines in the lower panels show our size cut of 0\farcs2. Blue points below the
size cut are excluded from our sample. The fraction of the blue points that are
excluded due to the small sizes is labeled in the lower panels for each \mstar\
bin (starting from $\logm=9$ and increasing with a width of 0.5 dex).
\label{fig:sample}}
\vspace{-0.2cm}
\end{figure*}

\subsection{Sample}
\label{data:sample}

We select SFGs with \mstar${\rm > 10^9 M_\odot}$, SSFR${\rm >10^{-1}
Gyr^{-1}}$, and $0.5<z<3$ in both fields to study \fclumpy. To ensure a clean
source detection with small photometric uncertainty in the F160W band, we also
require all galaxies to have $H_{F160W}<24.5$ AB. This apparent magnitude cut
only affects the mass completeness of our sample at $z>2$. As shown in the
lower panel of Figure \ref{fig:hmagcut}, a typical SFG (for example, a constant
star-forming model with age of 0.5 Gyr) with dust extinction $E(B-V)=0.15$ and
$\logm>9.0$ has an apparent F160W magnitude brighter than 24.5 AB at $0.5<z<2$.
At $z>2$, the mass completeness limit under this apparent magnitude cut
increases with redshift and reaches $\logm \sim 9.4$ at $z=3$. In our later
analyses, we still include galaxies with \mstar\ down to $\logm=9.0$ at $z>2$,
but remind readers that our lowest \mstar\ bin at $z>2$ is incomplete because
of the apparent magnitude cut.

The apparent magnitude cut of $H_{F160W}<24.5$ AB also ensures us a reliable
morphology and size measurements of our galaxies. In our study, the size
(semi-major axis, $r_e$ or SMA hereafter) and axial ratio ($q$) of each galaxy
are taken from \citet{vanderwel12}, who measured these parameters by running
GALFIT \citep{galfit} on the CANDELS F160W images.  \citet{vanderwel12} showed
that the random uncertainty of both $r_e$ and $q$ is $\sim$20\% at
$H_{F160W}=24.5$ AB, and quickly increases to about $\sim$50\% at
$H_{F160W}=25.5$ AB. The SFR--\mstar\ and size (SMA)--\mstar\ relations of our
sample are shown in Figure \ref{fig:sample}.


We also exclude galaxies whose sizes are less than 0\farcs2, because clumps
cannot be resolved in these marginally resolved or unresolved sources.
In the lower panel of Figure \ref{fig:sample}, we give the fraction of the
galaxies that are excluded because of their small size in each \mstar\ and
redshift bin.  If we assume that galaxies in each (redshift, \mstar) bin are
self-similar despite their different sizes, \fclumpy\ and the clump
contribution measured from the resolved galaxies in later sections are still
representative for the whole SFG population in the (redshift, \mstar) bin. On
the other hand, if we believe that there are physical reasons that make the
unresolved galaxies non-clumpy, we should scale down our the \fclumpy\ and the
clump contribution in our later analyses by the fraction of the unresolved
galaxies in each (redshift, \mstar) bin. 

The above two assumptions are two extremes that our \fclumpy\ can be
easily used to infer the \fclumpy\ of {\it all} SFGs regardless of their sizes.
The real situation, however, could be in between the two extremes. For example,
smaller (unresolved) galaxies may have intrinsically fewer clumps and lower
\fclumpy. In this case, the \fclumpy\ of unresolved galaxies cannot be simply
inferred from the \fclumpy\ of resolved galaxies. If that is true, current data
cannot address the \fclumpy\ of unresolved galaxies, observations with higher
spatial resolutions are needed.

To minimize the effect of dust extinction and clump blending, we only use
galaxies with axial ratio $q>0.5$. This $q$ criterion excludes some very
elongated clumpy galaxies, such as chain galaxies in \citet{elmegreen05} and
\citet{elmegreen07}. As shown by \citet{elmegreen05}, the axial ratio
distribution of chain galaxies plus clump-clusters is constant, as expected for
randomly oriented disks. \citet{ravindranath06}, however, found that the axial
ratio distribution of high-redshift Lyman Break Galaxies is skewed toward the
high-value end, against the scenario of randomly oriented disks. Although the
galaxy number distribution in our sample is skewed toward lower $q$ in the
high-redshift low-mass range (i.e., more red points than blue points above the
size cut at $\logm<10$ in the lower right panel of Figure \ref{fig:sample}),
galaxies with $q > 0.5$ (blue points) and $q \leq 0.5$ (red points) follow
almost the same SFR--\mstar\ and SMA--\mstar\ relations in Figure
\ref{fig:sample}. Therefore, we believe that, in general, the properties of the
clumps in $q > 0.5$ galaxies are likely to be representative of those in all
SFGs, regardless of their inclinations. Furthermore, excluding very elongated
galaxies reduces the rate of problematic detections by our clump finder, which
tends to over-deblend elongated galaxies.

After the above selection criteria, and further excluding galaxies that are not
covered by the ACS images, the final sample consists of 3239 galaxies.

\section{Detecting Discrete Star-Forming Regions}
\label{clfinder}

We begin our clump definition by searching for clumps among discrete
star-forming regions, believing that clumps occupy the bright end of the
luminosity distribution of the discrete star-forming regions. Before separating
clumps from ordinary star-forming regions (i.e., individual or blended HII
regions), we call all regions detected in this section ``blobs'' for
simplicity.

\subsection{Automated Star-Forming Region Finder}
\label{clfinder:code}

\begin{figure}[htbp]
\center{\includegraphics[scale=0.3, angle=0]{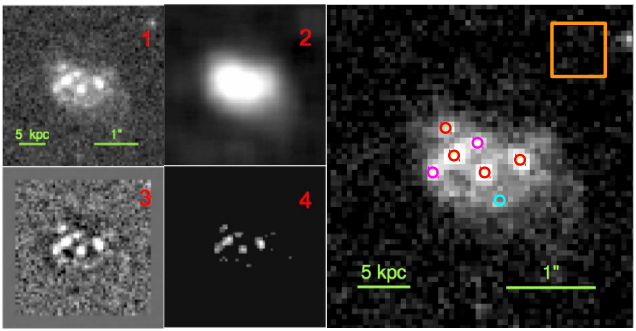}}
\caption[]{Illustration of the process of our blob finder. First, the original
image (panel 1) is smoothed. The smoothed image (panel 2) is then subtracted
from the original image to make a contrast image (panel 3). After low-S/N
pixels are masked out, blobs are detected from the filtered image (panel 4).
The final detected blobs (red and magenta) are shown in
the right panel. The orange box in the right panel shows the size of the
smoothing box (0\farcs6). The blob detection depends on the size of the
smoothing box. If the box size is reduced by half to 0\farcs3, only red blobs
are detected. And if the box size is doubled to 1\farcs2, a new blob (cyan)
will be added to the detection.
\label{fig:code}}
\vspace{-0.2cm}
\end{figure}

We design an automated blob finder to detect blobs from the galaxies in our
sample. The process of the blob finder is illustrated in Figure \ref{fig:code}.
We first cut a postage stamp for each galaxy from its clump detection image.
The size of the postage stamp is determined by the dilated segmentation area of
the source.  The process of ``dilation'' extends the SExtractor F160W
segmentation area generated in our source detection
\citep{ycguo13goodss,galametz13uds} to a proper size to include the outer wing
of the object below the SExtractor isophotal detection threshold
\citep[see][for details]{galametz13uds}. We then smooth the postage stamp
(Panel 1) through a boxcar filter with size of 10 pixels (0\farcs6) to
obtain a smoothed image (Panel 2).  Then, we subtract the smoothed image from
the original image to make a contrast image (Panel 3). The above steps are
similar to those used in calculating the ``Clumpiness (S)'' of the CAS system
of \citet{conselice03}.  We then measure the background fluctuation from the
contrast image after $3\sigma$-clipping. We then mask out (set value to 0) all
pixels below 2$\sigma$ of the background fluctuation to make a filtered image
(Panel 4), where blobs stand out in a zero background. We then run SExtractor
on the filtered image to detect sources, requiring a minimal detection area of
5 pixels to exclude spurious detections.  Each detected source is considered as
one blob. In the example of Figure \ref{fig:code}, the detected blobs are shown
by red symbols in the right panel. A comparison with the CANDELS visual
clumpiness \citep{kartaltepe14} shows our automated finder works well for
identifying discrete star-forming regions (see Appendix \ref{clfinder:visual}).

\begin{figure*}[htbp]
\hspace*{-0.1cm}
\vspace*{-0.1cm}
\includegraphics[height=2.8in,angle=0,clip]{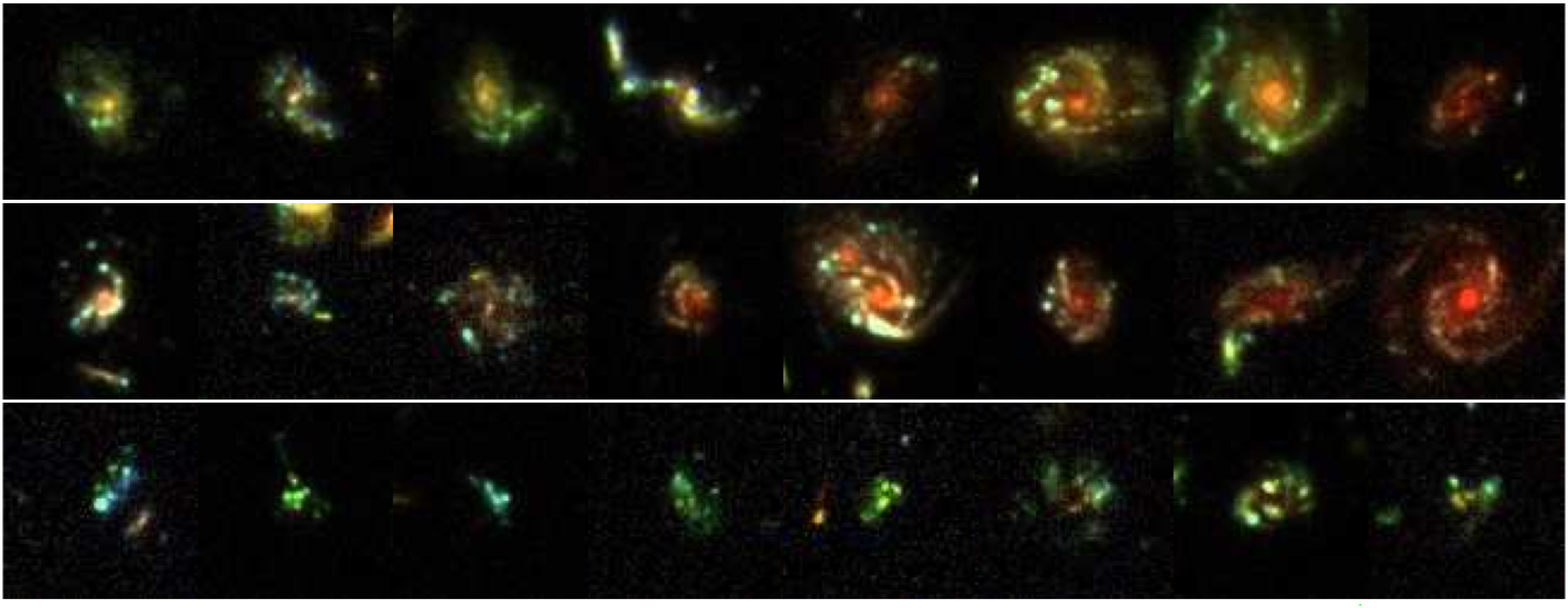} \\
\includegraphics[height=2.75in,angle=0,clip]{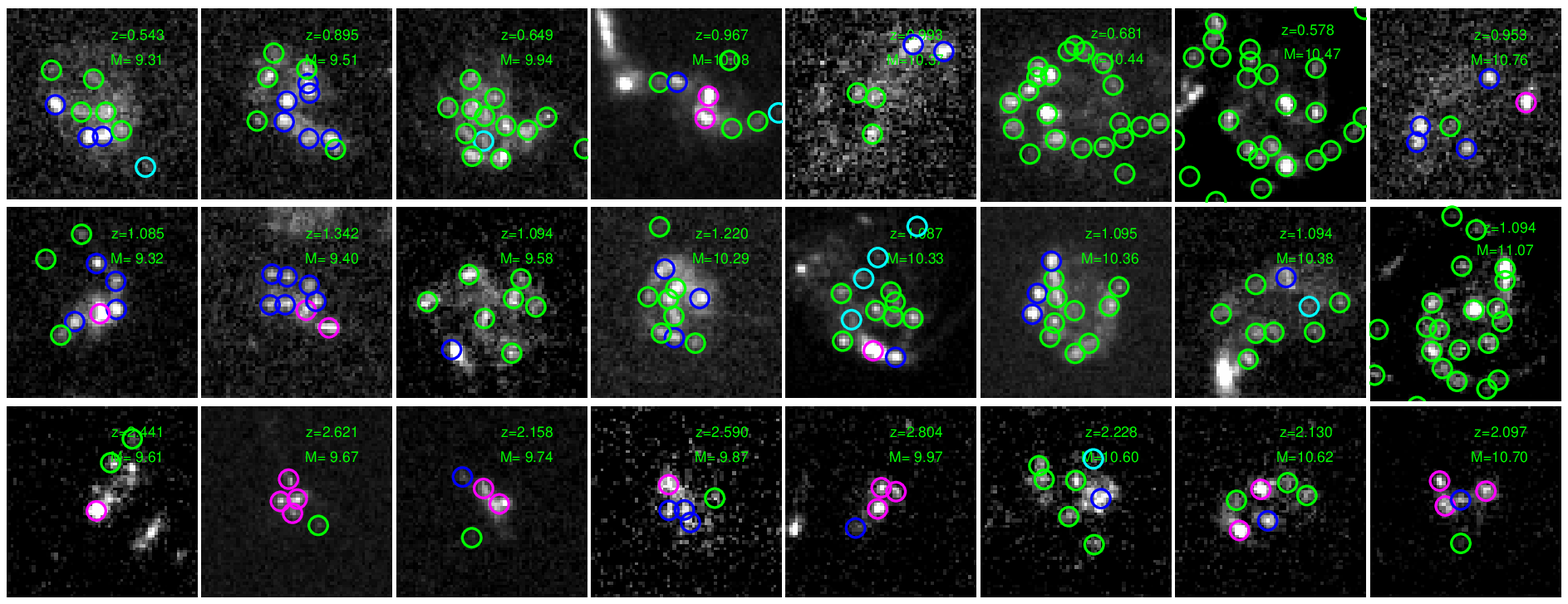}
\caption[]{Examples of visually clumpy galaxies and blobs detected by our
automated blob finder. The first three rows show the composite RGB images made
by the F435W, F606W, and F850LP images of the galaxies. The last three rows
show the same galaxies in the images used to detect blobs. The detected blobs
are shown by circles. The color of each circle shows the fractional luminosity (${\rm
FL} = L_{blob}/L_{galaxy}$) of the blob: magenta, ${\rm FL}>0.1$; blue,
$0.05<{\rm FL}<0.1$; green, $0.01<{\rm FL}<0.05$; and cyan, ${\rm FL}<0.01$.
The redshift and \mstar\ of each galaxy are labeled. For each row, the \mstar\
increases from the left to the right, while the redshift increases from the top
to the bottom row. In order to show as many as possible examples of blobs,
these galaxies are intentionally chosen to have very high clumpiness from the
CANDELS visual classification in the CANDELS/GOODS-S field. (see Appendix
\ref{clfinder:visual}).  Note that the image scales of the first three rows are
different from those of the last three rows.
\label{fig:showcase}}
\vspace{-0.2cm}
\end{figure*}

The blob detection depends on the size of the smoothing box. Our choice of 10
pixels (0\farcs6), shown by the orange box in the right panel of Figure
\ref{fig:code}, is the optimized one according to our later test of fake blobs
(Sec. \ref{clfinder:complete}) and comparison with the CANDELS visual
inspection (Appendix). As long as the smoothing length is significantly larger
than the typical size of blobs, blobs would stand out in the contrast image
(panel 3 of Figure \ref{fig:code}) and hence be detected. Since the smoothing
length of 0\farcs6 ($\sim$5 kpc at $0.5<z<3$) is significantly larger than the
typical size of blobs ($<1$ kpc), most of the UV-bright blobs should be able to
stand out in the contrast image unless their sizes are close to 5 kpc. If,
however, the smoothing length is too large, some noisy pixels may also be able
to stand out in the contrast image and hence be detected as a blob. We
demonstrate the effect of using different smoothing lengths in Figure
\ref{fig:code}. If we use 5 pixels (0\farcs3) to smooth the image, two obvious
blobs (magenta) would be missed.  On the other hand, if we use 20 pixels
(1\farcs2), a new blob (cyan) would be detected. This cyan blob, however, is
likely a spurious detection and would be excluded by our later clump definition
(Sec. \ref{definition}). More examples of identified blobs in clumpy galaxies
can be found in Figure \ref{fig:showcase}.

\begin{figure*}[htbp]
\includegraphics[height=3.5in,angle=0,clip]{./fig5a.ps}
\includegraphics[height=3.5in,angle=0,clip]{./fig5b.ps}
\includegraphics[height=3.5in,angle=0,clip]{./fig5c.ps}
\caption[]{Light profiles of the detected blobs in GOODS-S. Here we keep the
local background of the blobs, but we subtract it when measuring blob fluxes.
The blobs are divided into different bins based on their fractional luminosity
(${\rm FL}=L_{blob}/L_{galaxy}$), redshift, and \mstar\ of their host galaxies.
In each panel, each gray line shows the profile of one blob. The red solid line
shows the averaged profile of this bin. The red and blue dashed lines show the
1$\sigma$ and 2$\sigma$ ranges. The black solid line shows the light profile of
the corresponding PSF of the detection band plus the average background of the
blobs in this bin. The vertical yellow lines shows the aperture size used to
measure the blob fluxes.
\label{fig:clprofile}}
\vspace{-0.2cm}
\end{figure*}

\subsection{Detection and Measurement}
\label{clfinder:photo}

We detect blobs in different \hst/ACS bands based on the redshift of the
galaxies. The choice of the detection filter in GOODS-S is shown in the upper
panel of Figure \ref{fig:hmagcut}: F435W for galaxies at $0.5<z<1.0$, F606W for
galaxies at $1.0<z<2.0$, and F775W for galaxies at $2.0<z<3.0$. The purpose of
the choice is to detect blobs in the same rest-frame UV range, namely 2000 \AA
-- 2800 \AA, at different redshifts. For UDS, we use F814W to replace the F775W
for galaxies at $2.0<z<3.0$. There are, however, no \hst\ observations close to
F435W available in UDS. As a compromise, we use F606W to detect blobs for
galaxies at $0.5<z<1.0$ in UDS. We will discuss the systematic offsets
introduced by this band mis-match later.

Once a blob has been detected, we measure its flux in the detection band by
assuming it is a point source. The assumption is validated by the statistics of
the light profile of all detected blobs in Figure \ref{fig:clprofile}. The
average light profile of blobs with given fractional luminosity (${\rm FL} = L_{blob}/L_{galaxy}$) in a given redshift and host galaxy mass bin is very
well described by the light profile of the PSF of the detection band plus the
average background of the blobs. Here we assume the light profile at $r$$>$6
pixel (0\farcs36) is dominated by the background (``disk'' component) light.
The only exception happens for faint blobs (${\rm FL} < 0.1$) in
the lowest redshift bin ($0.5<z<1.0$), where the average blob profile is
broader than that of the PSF. The PSF profile, however, still lies within the
1$\sigma$ range of the blob profiles, implying that the blobs are only
marginally resolved. Overall, we conclude that the detected blobs are just
marginally, if at all, resolved and the assumption of a point source would not
introduce significant systemics in measuring the blob fluxes.

When measuring the flux of each blob, we first determine the background light
from the azimuthally averaged flux at $r$=6--10 pixels away from the blob
center, after masking out the central region ($r<4$ pixels) of all other blobs.
We then extrapolate the background flux to the center of the blob.  After
subtracting the background, we measure an aperture flux with radius $r$=3
pixels. This background-subtracted aperture flux is finally scaled up based on
the curve-of-growth of the corresponding PSF to obtain the total flux of the
point-like blobs.

We choose to subtract the local background of blobs, because we believe that
the blobs are ``embedded'' in the galaxies. Whether or not the local background
should be subtracted is still an open issue in clump studies
\citep[e.g.,][]{fs11b,ycguo12clump,wuyts12}. In fact, the background
subtraction is also a controversial issue for studying local star-forming
regions. It even affects our understanding of the basic physics of star
formation, e.g., the slope of the Kennicutt-Schmidt Law (see the comparison
between \citet{bigiel08} and \citet{gliu11}). If we do not subtract the local
background and scale up the total aperture flux within $r$=3 pixels according
to the PSF profile, the fluxes of our blobs will be systematically higher by a
factor of two. 

\subsection{Completeness of the Blob Finder}
\label{clfinder:complete}

\begin{figure*}[htbp]
\includegraphics[scale=0.3,angle=0,clip]{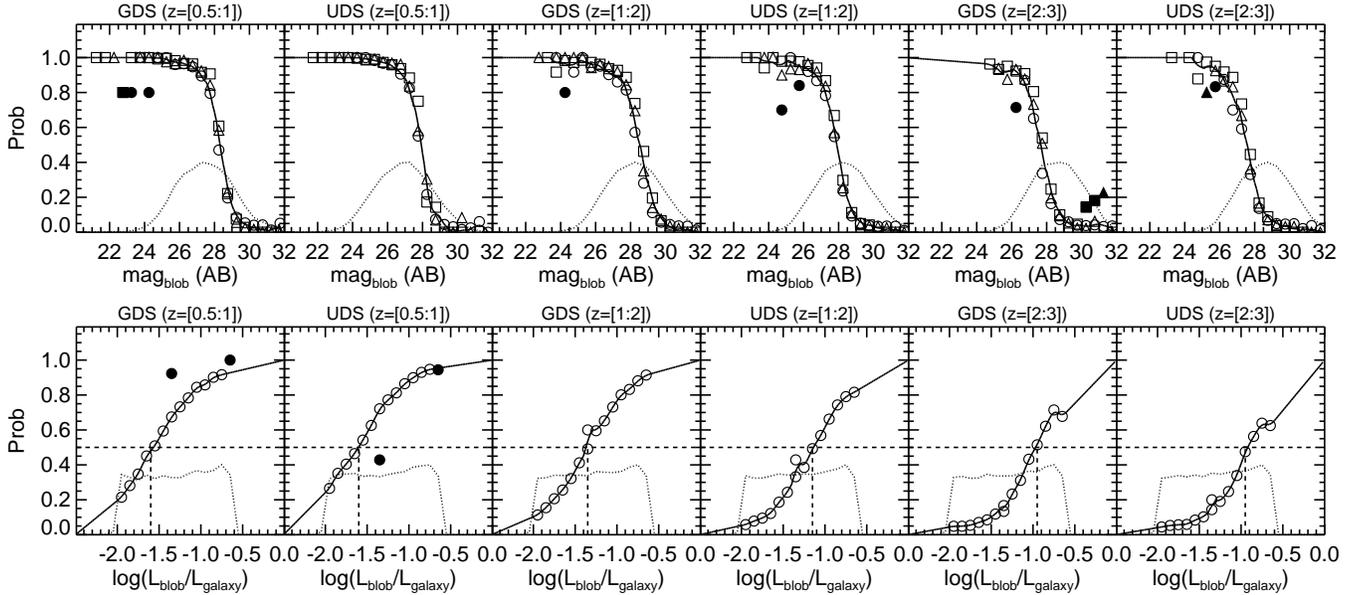}
\vspace{-0.2cm} 
\caption[]{Detection probability of fake blobs, namely, the successful rate of
recovering fake point sources, of our blob finder as a function of the
magnitude of fake blobs (upper panels) and the fractional luminosity of fake
blobs (lower panels). Detections in different fields and different redshifts
are shown in different panels (GDS: the GOODS-S field and UDS: the UDS field). 
In each panel, the dotted curve in the bottom shows the distribution of the
parameter of the fake blobs.  The solid curves across data points show the
interpolation to the data, with the filled data points being excluded. The
dashed horizontal lines in the lower panels show the detection probability of
50\%, while the dashed vertical lines show the corresponding fractional
luminosity of the 50\% detection probability.
\label{fig:complete}}
\vspace{-0.2cm}
\end{figure*}


We evaluate the completeness of our blob finder by recovering fake blobs. For
each galaxy in our sample, regardless of whether it contains detected blobs, we
insert one fake blob into its image in the detection band and re-run our blob
finder on it. We use point sources to mimic the blobs. This simplification is
validated by the fact that the light profile of blobs can be well described by
the PSF of the detection bands (Figure \ref{fig:clprofile}). The fluxes of fake
blobs are randomly selected from a uniform distribution between 1\% and 20\% of
the flux of their galaxies. The fake blobs are only added into the segmentation
areas of the galaxies. For each galaxy, we repeat the process 30 times to
improve the statistics. Comparing with the method of adding arbitrary numbers
of blobs to fake model galaxies (e.g., \sersic\ models), our method largely
preserves the distributions of the size, magnitude, surface brightness profile,
and blob crowdedness of real galaxies, which are all important to the blob
detection probability.

The detection probability, i.e., the successful rate of recovering fake blobs,
depends on the properties of both galaxies and blobs. More specifically, it
depends on redshift ($z$), the magnitude of galaxies ($mag_g$), the size of
galaxies ($r_e$), the magnitude of blobs ($mag_b$), the location of blobs (the
distance to the center of the galaxies, $d_b$), and the number of blobs in the
galaxies ($n_b$). 
For each of the real blobs, we assign a detection probability to it based on
its values of the above parameters, $P(z, mag_g, r_e, mag_b, d_b, n_b)$, if we
have at least five detected fake blobs in the $(z, mag_g, r_e, mag_b, d_b,
n_b)$ bin. Otherwise, we determine its probability by interpolating the
marginalized detection probability as a function of the FL of the blobs (the
second row of Figure \ref{fig:complete}). In fact, using the
probability--$mag_b$ relation (the first row of Figure \ref{fig:complete}) also
provides a good approximation for blobs in the under-sampled bins, but using the
probability--FL relation makes our later analyses easy because we are measuring
the FLF instead of the absolute luminosity function. Only $\lesssim$10\% of our
blobs fall in the under-sampled $(z, mag_g, r_e, mag_b, d_b, n_b)$ bins. Using
the interpolated marginalized detection probability would not affect our later
results. 

In order to avoid possible contamination from bulges, which usually stand out
in the filtered images (Panel 3 of Figure \ref{fig:code} and hence almost
always are detected as blobs, we also exclude blobs that are within $d_b<0.5
\times r_e$. For example, we only count five blobs in the galaxy in Figure
\ref{fig:code}. We also exclude blobs that are beyond $d_b>8 \times r_e$ (if
the size of the postage stamp of a galaxy is larger than $8 \times r_e$), in
order to reduce the impact of nearby small satellite galaxies.

We also measure the fluxes of the fake blobs using the method described in Sec.
\ref{clfinder:photo} and compare them with the input values. In general, the
measured and input values show good agreement. There is, however, a mild trend
that the fluxes are overestimated as the galactocentric distance of the blobs
($d_b$) decreases, with the maximum overestimation of $\sim$30\% for blobs at
$0.5\times r_e$. We fit the overestimation--$d_b$ relation and scale down the
flux of each real blob based on its $d_b$.

\begin{figure*}[htbp]
\includegraphics[scale=0.4,angle=0,clip]{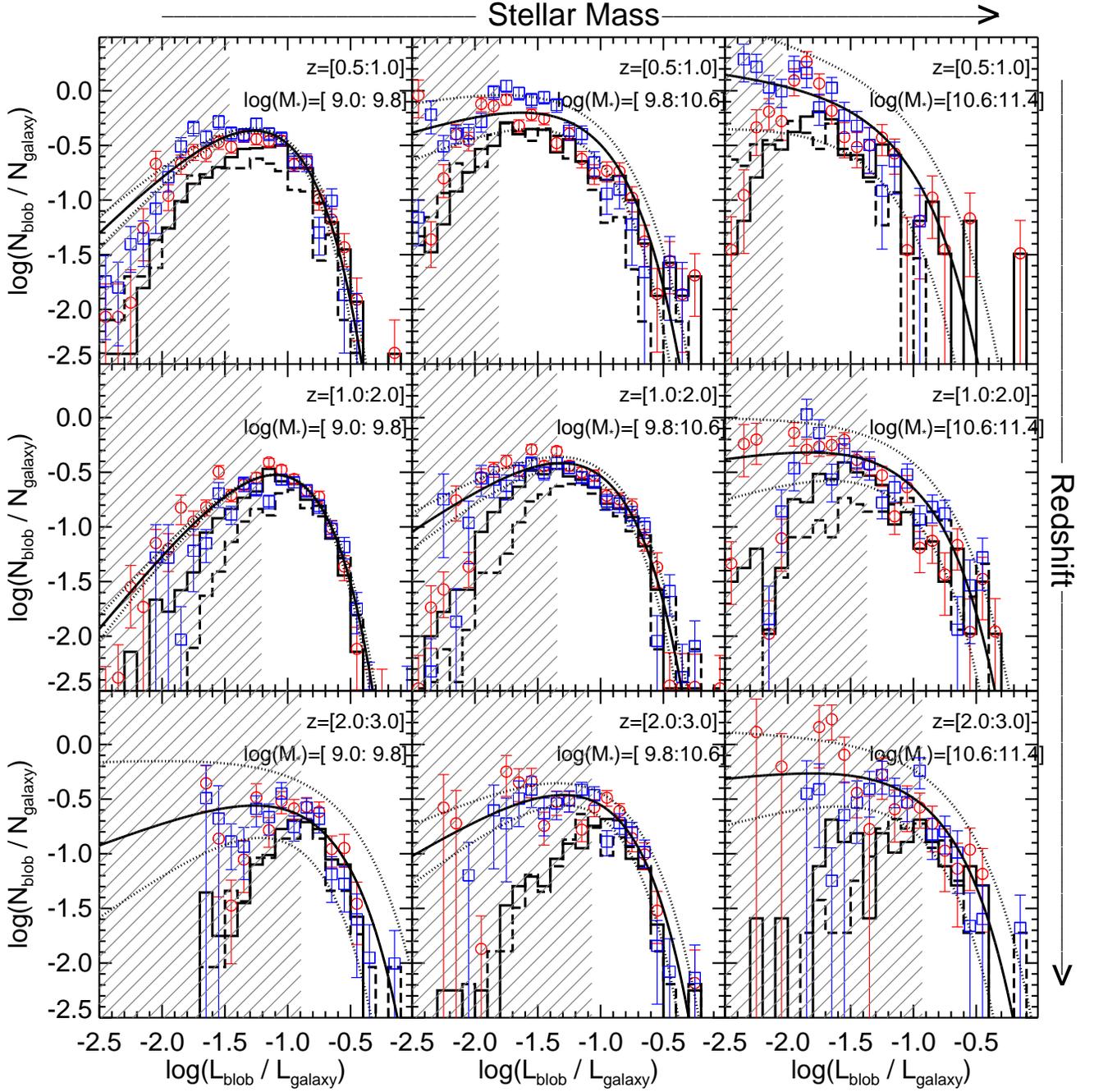}
\caption[]{Fractional luminosity functions of blobs. Each panel shows the
average number of blobs per galaxy, as a function of the fractional luminosity
of the blobs, in galaxies within a given redshift and \mstar\ bin. Solid
(GOODS-S) and dashed (UDS) histograms show the results without being corrected
for the blob detection incompleteness.  Red (GOODS-S) and blue (UDS) symbols
show the results after the incompleteness correction. Error bars are derived
from the Poisson error of the blob number counts. The shaded area in each panel
shows the region where the blob detection incompleteness is larger than 50\%
(see the dashed vertical lines in the second row of Figure \ref{fig:complete} for an example of how the 50\% threshold is determined, but note that each panel of Figure \ref{fig:complete} includes galaxies with all \mstar, while galaxies are separated into different \mstar\ bins in this figure). The solid and
dashed black curves in each panel show the best-fit Schechter Function and its
confidence interval for the combined GOODS-S and UDS fractional luminosity
functions.
\label{fig:clfuv}}
\vspace{-0.2cm}
\end{figure*}

\begin{figure*}[htbp]
\includegraphics[height=3.5in,angle=0,clip]{./fig8a.ps}
\includegraphics[height=3.5in,angle=0,clip]{./fig8b_new.ps}
\caption[]{{\it Left}: Best-fit Schechter functions to the fractional
luminosity functions of blobs in all redshift and galaxy \mstar\ bins. The
shaded area shows the region where the blob detection incompleteness is larger
than 50\% for galaxies at $2<z<3$, while the double-shaded area shows the same
region at $1<z<2$. The 50\% incomplete region for galaxies at $0.5<z<1$ is
about $log(L_{blob}/L_{galaxy}) = -1.5$. {\it Right}: Best-fit $\Phi^*$ and
${\mathscr L}^*$ of the Schechter functions of all redshift and galaxy \mstar\ bins.  Red, blue, and black symbols are for galaxies at $0.5<z<1$, $1<z<2$, and $2<z<3$,
respectively.
\label{fig:clfuv_fit}}
\end{figure*}

\section{Fractional Luminosity Function of Blobs}
\label{fuv}


Now, we measure the FLF of the blobs, taking into account the detection
incompleteness.  Since our fake blobs only have the fractional fluxes down to
$L_{blob}/L_{galaxy}=0.01$, we extrapolate the detection probability for
fainter blobs using their fractional luminosity. It is important to note that
the incompleteness estimated in Sec. \ref{clfinder:complete} only tells us the
fraction of blobs that are missed by our blob finder. It does not tell us from
which galaxies, ``blobby'' or ``non-blobby'', they are missed.  Some
``non-blobby'' galaxies may actually contain a few blobs, which are somehow
missed in our detection. Since these missed blobs are taken into account in the
incompleteness, their host galaxies, which are mis-classified as ``non-blobby'',
should also be taken into account when we measure the FLF. Therefore, our FLF
and later UV light and SFR contributions from blobs (or clumps) are measured
for all SFGs rather than just for the galaxies with detected blobs (or clumps).

The FLFs of the GOODS-S and UDS fields, both before and after the
incompleteness corrections, are shown in Figure \ref{fig:clfuv}. The results at
$1<z<2$ are very encouraging, demonstrating that our fake blob test correctly
evaluates the incompleteness of our blob detection. In this redshift bin, both
GOODS-S and UDS fields select blobs from the \hst\ F606W band, but the
depths of their F606W images are different.  The GOODS-S image is about two
times deeper (in terms of exposure time) than the UDS one. As a result, the
uncorrected FLF (black histograms in the figure) of GOODS-S is about 1.5--2
times higher than that of UDS for blobs with $L_{blob}/L_{galaxy}<0.1$.  After
correcting the incompleteness, both functions (symbols with error bars) of
GOODS-S and UDS show excellent agreement in all three \mstar\ ranges. This
result indicates that after the correction, our results are largely unaffected
by the varying observation depth from field to field.

At $0.5<z<1$, 
due to the lack of F435W images in the UDS field, we must use the CANDELS
parallel F606W image to detect blobs. At this redshift range, F606W samples the
rest-frame U-band, while F435W samples the rest-frame 2500\AA. As found by
\citet{wuyts12}, the UV luminosity contribution of blobs decreases as the
detection bands shift from blue to red. If we extrapolate the UV luminosity
contribution of star-forming regions detected in different bands of
\citet{wuyts12} to the rest-frame 2500\AA, the difference of blob luminosity
contribution between 2500\AA\ and U-band is about a factor of 1.7.  After being
scaled up, the UDS FLF matches the GOODS-S FLF very well at $0.5<z<1$ in all
\mstar\ bins (the top panels of Figure \ref{fig:clfuv}).

In the highest redshift bin, $2<z<3$, the incompleteness corrected results of
the two fields also show agreement, but with larger uncertainties for faint
blobs ($L_{blob}/L_{galaxy}<0.03$), which are hard to detect at such high
redshifts. With a very small number of detected faint blobs, our incompleteness
correction method has difficulty to properly recover the real blob numbers. We
note that, however, our later analyses use little information from these
high-redshift faint blobs.

It is important to note that the faint end of each incompleteness-corrected FLF
in Figure \ref{fig:clfuv} should be treated with caution. In most panels, the
FLF decreases in the faint end, suggesting that the incompleteness is somehow
not properly corrected in the faint end, although our correction method shows
encouraging results in the bright and intermediate regions. 
Some small and faint blobs would be missed by our blob finder, because their
sizes do not satisfy our minimal area requirement of 5 pixels. Such blobs may
not be properly taken into account in our fake blob simulations, which results
in an underestimate of the incompleteness.

In Figure \ref{fig:clfuv}, we shadow the regions where the incompleteness is
larger than 50\%, i.e., the marginalized probability of detecting a fake
blob as a function of the FL of blob is less than 50\%. The second row of
Figure \ref{fig:complete} shows an example of how this 50\% threshold is
determined. It should be noted that in Figure \ref{fig:complete}, we do not
separate galaxies into different \mstar\ bins, but we do so in Figure
\ref{fig:clfuv}. Therefore, the 50\% thresholds in Figure \ref{fig:clfuv} are
slightly different from those in Figure \ref{fig:complete}. Also, we use the
average threshold of GOODS-S and UDS in each panel of Figure \ref{fig:clfuv}.
In the shaded regions, the shape of the FLF depends on the accuracy of our
incompleteness correction method more than on the number of detected blobs.


To study the evolution of the FLFs with redshift and \mstar, we fit a Schechter
Function \citep{schechter76} to each FLF in Figure \ref{fig:clfuv}:
\begin{equation}
n({\mathscr L})d{\mathscr L} = \Phi^* \times ({\mathscr L}/{\mathscr L}^*)^\alpha \times e^{-({\mathscr L}/{\mathscr L}^*)} d{\mathscr L}, 
\label{eq:clfuv}
\end{equation}

where ${\mathscr L} = L_{blob} / L_{galaxy}$. The best-fit functions and their
parameters, $\Phi^*$ and ${\mathscr L}^*$, are shown in Figure
\ref{fig:clfuv_fit}. We do not show $\alpha$ because it strongly depends
on the very faint end of the fractional luminosity function (e.g.,
$L_{blob}/L_{galaxy} \sim 0.01$), where our blob detection completeness is very
low (only 10--20\%). The dependences of $\Phi^*$ and ${\mathscr L}^*$ on the
very faint end are weaker than that of $\alpha$.
For the
ranges of $0.5<z<1$ and $1<z<2$, we fit the functions down to the faint
luminosity of $log({\mathscr L})=-2.5$, while in the highest redshift range,
$2<z<3$, we only fit the functions down to $log({\mathscr L})=-2.0$ due to the
large error bars and missing data in some luminosity bins (e.g., blobs around
$L_{blob}/L_{galaxy} \sim -2.0$ in the least massive bin in this redshift in
Figure \ref{fig:clfuv}). We also overplot the best-fit functions and their
uncertainty ranges in Figure \ref{fig:clfuv}. It is important to note that the
choice of the Schechter Function is empirical and not driven by any physical
reasons. In fact, a truncated power-law is usually used to study the
bright end of the luminosity functions of nearby HII regions
\citep[e.g.,][]{scoville01,gliu13}. We choose the Schechter Function because it
fits both the bright and faint ends.

The trends of the best-fit Schechter parameters can be clearly seen from
the right panels of Figure \ref{fig:clfuv_fit}.  For galaxies with a given
\mstar\, the characteristic fractional luminosity (${\mathscr L}^*$), namely
the characteristic blob contribution of the UV luminosity of their galaxies,
increases with redshift, while the number of the characteristic blobs per
galaxy ($\Phi^*$) decreases with redshift. This result shows that the lower the
redshift, the fainter (in terms of UV light contribution) the blobs are as well
as the larger their numbers are. Although the shift of the FLFs toward the
bright end from low to high redshifts could be physical and suggest a
transition of the star formation mode with redshift and \mstar\ \citep[e.g., as
indicated by][]{mandelker14}, it is more likely due to an observational effect:
the blending of blobs. The faint blobs are hard to detect individually at high
redshift as well as in low-mass galaxies due to the low spatial resolution in
physical length at high redshifts and the small size of the galaxies. If
detected blended, they will shift the FLF toward the bright side and suppress
the number of the faint blobs, resulting in a decline of the number of the
faint blobs toward high redshift and low-mass galaxies as seen in Figure
\ref{fig:clfuv_fit}.

\begin{figure*}[htbp]
\includegraphics[scale=0.4,angle=0,clip]{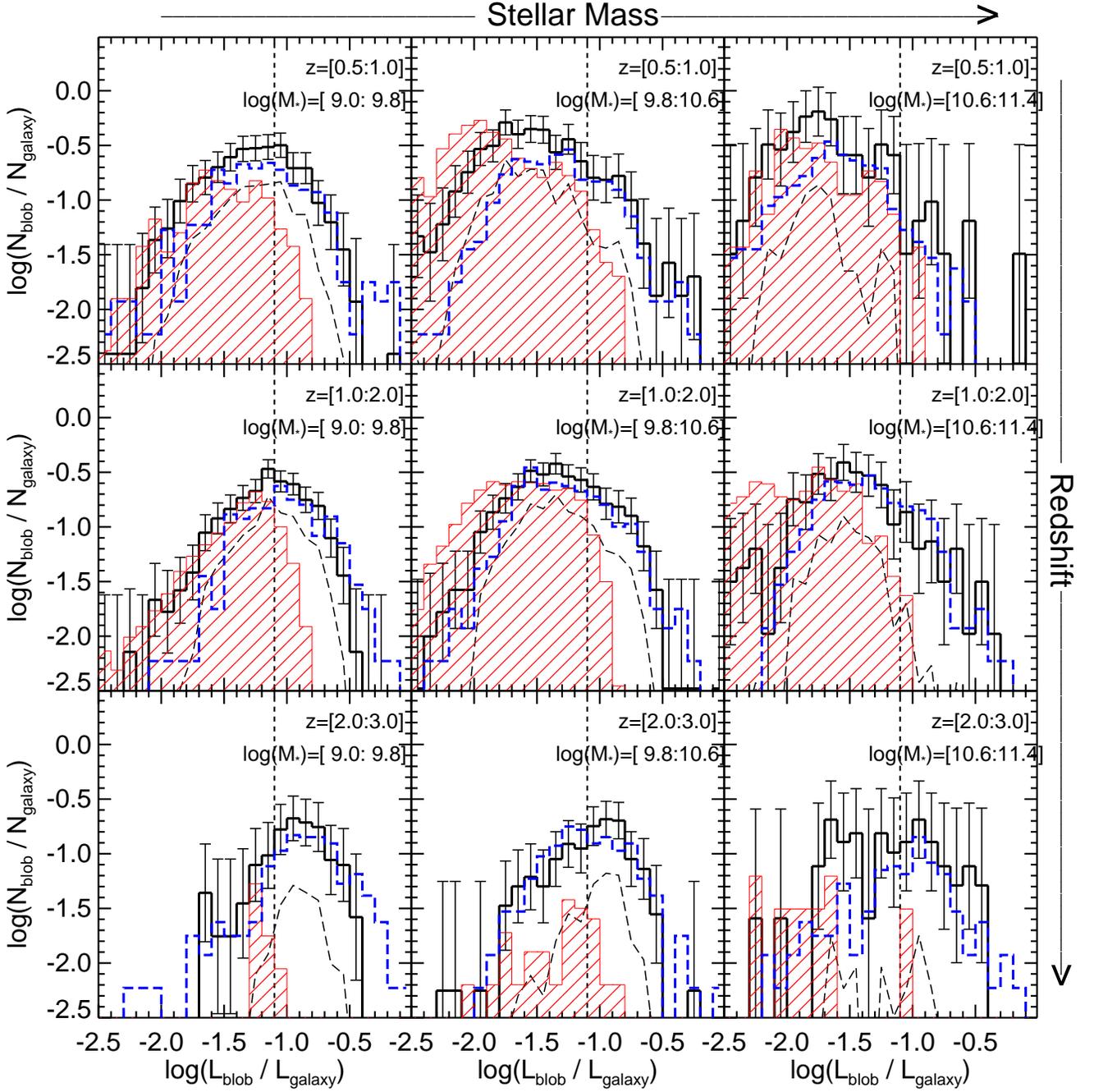}
\caption[]{Definition of clumps. In each panel, the fractional luminosity
function of GOODS-S (not corrected for the detection incompleteness) is shown
by the black histogram with error bars from the Poisson error. The black dashed
curve shows the lower 3$\sigma$ level of the error bars. The red shaded region
shows the fractional luminosity function of blobs detected from the fake
redshifted M101 galaxies. The vertical dashed line shows our definition of
clumps: blobs brighter than the line are defined as clumps. The blue dashed
histogram shows the fractional luminosity function of blobs detected in the
redshifted fiducial galaxies ($9.8<\logm<10.6$ and $0.5<z<1.0$, see Sec.
\ref{definition} for details).
\label{fig:compm101}}
\vspace{-0.2cm}
\end{figure*}

\section{A Physical Definition of Clumps}
\label{definition}

The issue of the blending of blobs also raises a question: are these
small blobs simply blended star-forming regions similar to those seen in nearby
disk or spiral galaxies? This reaffirms the problem faced by any clump
definition based on the appearance of galaxies: the natures of thus defined
clumps change with the redshift and size of galaxies due to observational
effects.

In order to understand to what extent our detected blobs can be statistically
described by the counterparts of local star-forming regions, we shift a grand
design spiral galaxy, M101 (NGC5457), to the redshift of each galaxy in our
sample and detect blobs from the redshifted images. We use the SDSS $u$-band
image for the test. At the distance of M101 (6.4 Mpc), the spatial resolution
of the SDSS image (1\farcs4) is equivalent to about 40 pc, sufficient to
resolve large HII regions. For each galaxy in our sample, we also shrink the
physical effective radius of M101 to match the effective radius of the galaxy.
We re-bin and smooth the SDSS images to match, in units of kpc, the pixel size
and spatial resolution of our \hst/ACS blob detection images.  We then re-scale
the total flux, in units of Analogue-to-Digital Unit (ADU), of the re-binned
and smoothed M101 image to match the total flux of each of our sample galaxies
in the blob detection band.  Therefore, the redshifted M101s are matched to the
redshift, size, and apparent surface brightness of each galaxy in our sample.
We finally add a fake background fluctuation, whose 1$\sigma$ level is equal to
that of our \hst/ACS detection images, to the re-scaled M101 image. In this
paper, we do not follow the rigid steps of redshifting a galaxy, such as
determining the morphological K-correction and cosmological dimming
\citep[e.g.,][]{barden08}, because our purpose is not to study how galaxies
with the M101 spectral type and luminosity look at higher redshifts. Instead,
our purpose is to study how galaxies with the M101 {\it appearance} look at higher
redshifts. The scaling of the flux of M101 provides a reasonable shortcut for
us \citep[see][for a detailed description of similar simulation
tests]{conselice03}.

We run our blob finder on the redshifted M101 images and measure the FLFs of
the detected blobs in each redshift and \mstar\ bin. The results are shown in
Figure \ref{fig:compm101}, overplotted with the observed GOODS-S FLFS, both
uncorrected for the detection incompleteness. The figure shows clearly that at
$z\leq2$, the faint end of the observed FLF in each \mstar\ bin can be well
explained by that of the redshifted M101s. This indicates that the faint blobs
detected by our automated finder are actually not statistically different from
the redshifted local HII regions, once the local galaxies are matched to the
size of the high-redshift galaxies. These faint blobs should be excluded from
our definition of clumps. The comparison at $z>2$ is not as conclusive as that
at $z\leq2$ due to the large Poisson error bars of the observed functions. But
still, we see a hint that the redshifted local HII regions can explain a large
fraction of the faint end of the observed functions, which suggests that
similar observational effects of blurred and blended local HII regions are also
present at $z>2$.

In this paper, we define clumps as blobs whose fractional luminosities are
significantly higher than that of redshifted star-forming regions of 
nearby large spiral galaxies.
Particularly, we choose a threshold where the observed FLF is $\sim 3 \sigma$
higher than the FLF of the redshifted M101s. This threshold (where the dashed
curves cross the red shaded histograms in Figure \ref{fig:compm101}) changes
slightly among different (redshift, \mstar) bins. For simplicity, we choose the
threshold as $L_{blob}/L_{galaxy} = 0.08$ and thus define clumps as blobs whose
UV luminosity is brighter than 8\% of the total UV luminosity of the galaxies
(as shown by the vertical dashed lines in Figure \ref{fig:compm101}). This
definition of clumps takes into account the observational effects due to the
sensitivity and resolution as well as the change of size of galaxies with
redshift and \mstar. Therefore, it defines clumps in a more physical way than
the appearance of galaxies and can be easily applied to galaxies at different
redshifts regardless of the observational effects.

We also test the 3$\sigma$ threshold of our clump definition by using other
nearby spiral galaxies and find that the threshold is only mildly changed. We
repeat the above test by redshifting the GALEX NUV (spatial resolution of
5\farcs0) images of M83 and M33. The physical resolution is $\sim$100 pc and
$\sim$23 pc at the distance of M83 (4.61 Mpc) and M33 ($\sim$0.9 Mpc). For M83,
the 3$\sigma$ thresholds in all (redshift, \mstar) bins are quite close to
those in the M101 test with a different of at most 0.1 dex, except for in the
$z$=0.5--1.0 and $\logm<9.8$ bin, where the M83 threshold is 0.5 dex smaller
than the M101 threshold. For M33, the 3$\sigma$ thresholds are systematically
smaller than those of the M101 test by 0.2--0.3 dex in almost all (redshift,
\mstar) bins. In our later analyses, we keep using the value of
$L_{blob}/L_{galaxy}$=8\% (the vertical dashed lines in Figure
\ref{fig:compm101}) that is derived from the M101 test as the default
definition. We will also discuss how \fclumpy\ changes if we use an aggressive
definition of $L_{blob}/L_{galaxy}=0.05$ or a conservative definition of
$L_{blob}/L_{galaxy}=0.1$.

Our clump definition uses the HII regions of nearby large spiral galaxies
as the null hypothesis and rejects it once the event of a blob with $>$8\% UV
fractional luminosity happens. This definition is appropriate and necessary to
exclude ``non-clumpy'' galaxies, because the purpose of this paper is to carry
out a statistical census of clumpy galaxies. It is, however, important to note
that using some nearby galaxies as the null hypothesis does not mean all local
galaxies are ``non-clumpy''. In fact, \citet{elmegreen09b} carried out similar
tests of redshifting local galaxies and found that clumpy galaxies at
intermediate to high redshifts resemble local dwarf irregulars in terms of
morphology, number of clumps, and relative clump brightness. Therefore, a large
fraction of local low-mass galaxies also contain clumps. Our later result
(Figure \ref{fig:fclumpy}) also confirms this point. The purpose of this paper
is not to distinguish high-redshift clumps from local clumps. It is to
distinguish clumps from non-clumps (i.e., small blobs and small HII regions).
The redshifted local dwarf irregulars cannot serve as a null hypothesis to
reject ``non-clumpy'' galaxies in statistics, although they are excellent
high-resolution counterparts to study the physical properties of high-redshift
clumpy galaxies \citep[e.g.,][]{elmegreen09b}.

\section{Fraction of Clumpy Galaxies}
\label{fclumpy}

\subsection{Clumpy Fraction}
\label{fclumpy:cut}

\begin{figure*}[htbp]
\includegraphics[scale=0.43,angle=0,clip]{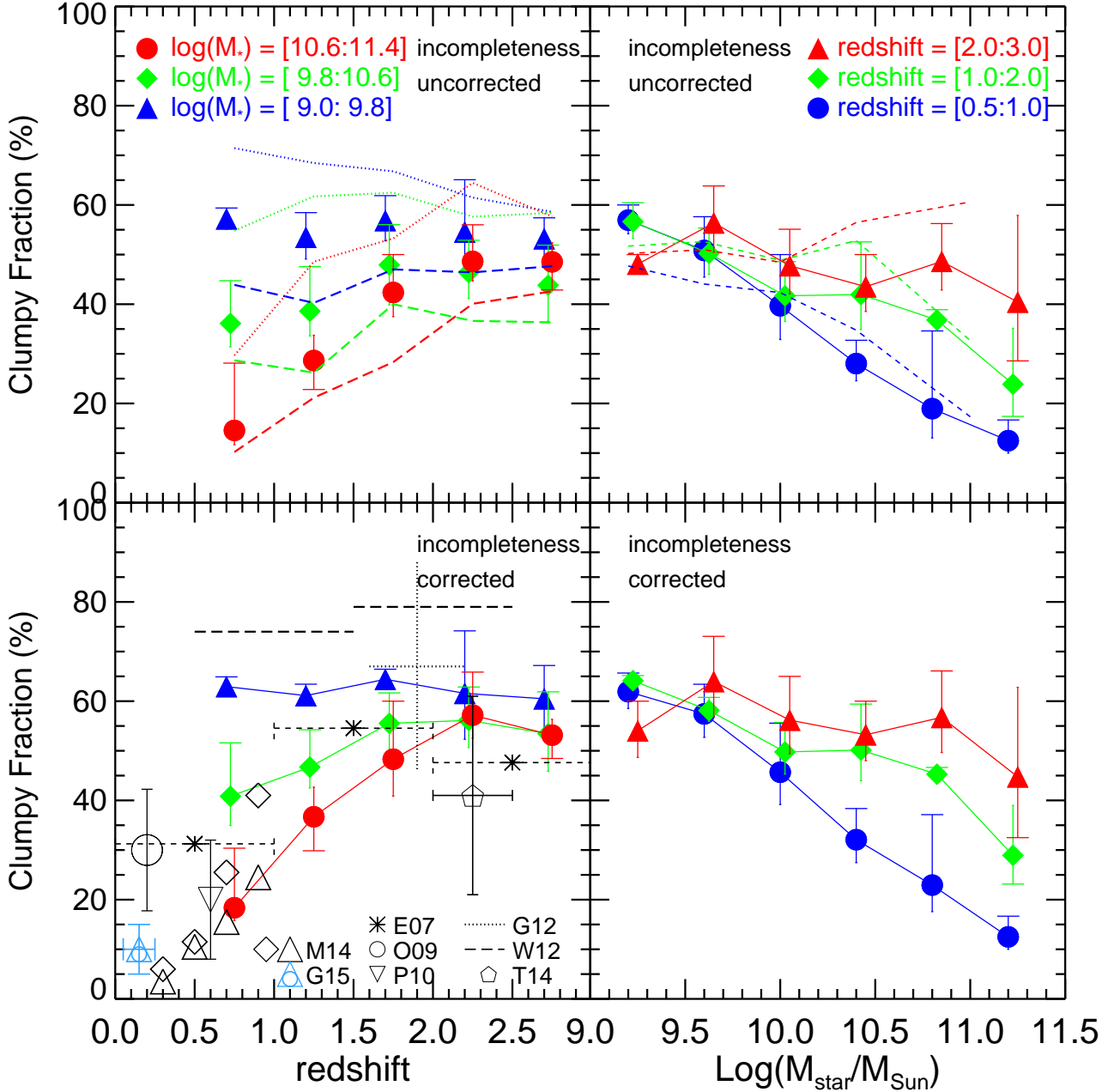}
\caption[]{Fraction of star-forming galaxies with at least one off-center UV
clump in different redshift and \mstar\ bins. The {\it upper} panels show the
results without correcting for the detection incompleteness, while the {\it
lower} panels show the results with correcting for the incompleteness through
Eq. \ref{eq:newfclumpy}. Each colored point is the error-weighted average of
the GOODS-S and UDS results. The hats of the upper and lower error bars of each
data point have different lengths: the longer hat shows the fraction of
GOODS-S, while the shorter one shows that of UDS. The errors of GOODS-S and UDS
fractions are not shown, but the relative errors between the two fields can be
inferred from the distances of each data point to the two hats of its error
bar. In the {\it upper left} panel, dashed and dotted lines show \fclumpy\
under an aggressive ($L_{blob}/L_{galaxy}=0.05$) and a conservative
($L_{blob}/L_{galaxy}=0.1$) clump definitions, respectively. The color of each
dashed or dotted line matches the color of the symbols to show its \mstar\
range. In the {\it upper right} panel, the dashed lines, also color-matched to
the symbols, show \fclumpy\ measured through comparing real galaxies with
redshifted fiducial galaxies to take into account the clump/blob blending
effects (see Sec. \ref{clumpfuv:vsms} for details). In the {\it lower left}
panel, several measurements of \fclumpy\ from other studies are also plotted.
The summary of the previous results is given in Table \ref{tb:fclumpy}.  
\label{fig:fclumpy}}
\end{figure*}

One of the main results of this paper --- the fraction of clumpy galaxies among
SFGs ($f_{clumpy}$) in a given \mstar\ and redshift bin --- is shown in Figure
\ref{fig:fclumpy}. Here clumpy galaxies are defined as galaxies that contain at
least one off-center ($d_b > 0.5 r_e$) clump as defined in Sec.
\ref{definition}. We measure the fraction and its uncertainty (Poisson errors
from number counts) separately for GOODS-S and UDS. Each color point in the
figure is the error-weighted average of the GOODS-S and UDS results. We also
show the fractions of the two fields as the hats of the error bar of each data
point.  Therefore, the error bars in the figure reflect the field variance
instead of the statistical uncertainty. The errors of the GOODS-S and UDS
fractions are not shown in the figure, but their relative strength can be
inferred from the distance of each data point to the two hats of its error bar.

The redshift evolution of $f_{clumpy}$ changes with \mstar\ of the galaxies
(the upper left panel of Figure \ref{fig:fclumpy}).  Low-mass galaxies
($\logm<9.8$) keep an almost constant \fclumpy\ around 55\%.
For intermediate-mass galaxies ($9.8<\logm<10.6$), $f_{clumpy}$ remains almost
constant around 45\% from z $\sim$3 to z$\sim$1.5, and then gradually drops to
$\sim$30\% at $z\sim0.5$. For massive galaxies ($10.6<\logm<11.4$),
$f_{clumpy}$ also keeps a constant of $\sim$50\% from z$\sim$3 to z$\sim$2, but
then quickly drops to $\sim$15\% at z$\sim$0.5. 

We also show \fclumpy\ under an aggressive ($L_{blob}/L_{galaxy}=0.05$) and a
conservative clump definition ($L_{blob}/L_{galaxy}=0.1$) in the {\it upper left} panel of Figure \ref{fig:fclumpy}. The
general trend of the \fclumpy--redshift relation of each mass range is not
significantly affected by the different definitions. The normalization of the
relations, however, is scaled up (down) by a factor of 1.2--1.3 for the
aggressive (conservative) definition.

The dependence of $f_{clumpy}$ on \mstar\ changes with redshift as well (the
top right panel of Figure \ref{fig:fclumpy}). In general, \fclumpy\ decreases
with \mstar\ in all redshift bins, but the slope of the trend depends on the
redshift. The lower the redshift, the steeper the slope (i.e., the faster
\fclumpy\ decreases with \mstar).

It is important to note that the above trends (top panels of Figure
\ref{fig:fclumpy}) are based on the direct number count of clumps without
taking into account the incompleteness of our clump detection. As shown in
Figure \ref{fig:complete} and \ref{fig:clfuv}, although the completeness is relatively high for our
clumps (i.e., blobs with high ${\rm L_{blob}/L_{galaxy}}$), it is still not 
unity. Therefore, we may underestimate $f_{clumpy}$ because of the missing
clumps. To correct for the incompleteness, we calculate a new $f_{clumpy}$
using the following formula, assuming the undetected clumps are randomly
distributed in the galaxies in our sample:
\begin{eqnarray}
f^{new}_{clumpy} = f^{old}_{clumpy} & + & \frac{1}{n_c} (\frac{1}{\it X}-1) (f^{old}_{clumpy}) \nonumber \\ 
 & - & \frac{1}{n_c} (\frac{1}{\it X}-1) (f^{old}_{clumpy})^2 
\label{eq:newfclumpy}
\end{eqnarray}
where $f^{old}_{clumpy}$ and $f^{new}_{clumpy}$ are the clumpy fractions before
and after the incompleteness correction is applied, $\it X$ the clump detection
completeness, and $n_c$ the average number of clumps in each clumpy
galaxy. The second term on the right hand side takes into account the
contribution of undetected clumps, while the third term takes into account
the fact that some undetected clumps may be in a galaxy that has already been
classified as clumpy, in which case the number of clumpy galaxies should not be
increased.

\begin{figure}[htbp]
\includegraphics[scale=0.45,angle=0,clip]{./fig11_new.ps}

\caption[]{Evolution of the fraction of galaxies with off-center UV clumps as a
function of redshift. Color symbols with error bars are identical to those in
the lower left panel of Figure \ref{fig:fclumpy}. 
The solid black curve shows the fraction of massive disks that are unstable. It
is derived by combining the prediction of disk instability in massive disk of
the two-component (gas + star) fiducial model of \citet{cacciato12} and the
kinematic measurement uncertainty of \citet{kassin12}. See the text for
details.  The dashed black lines are the minor merger rate of \citet{lotz11},
scaled by a merger observability timescale of 1.5, 2, and 2.5 Gyrs (from bottom
to top). The dotted black lines are the major merger rate of \citet{lotz11},
scaled by a merger observability timescale of 1, 2, and 3 Gyrs (from bottom to
top). The dotted-dashed line is the wet major merger fraction of \citet{ls13}.

\label{fig:fclumpy_merger}}
\vspace*{-0.1cm}
\end{figure}

\begin{table*}
\begin{center}
\caption[]{Summary of Papers and Samples Used for Clumpy Fraction Comparison
\label{tb:fclumpy}}
    \begin{tabular}{ | l | l | c | c | c | l |}
    \hline
    \hline
    Paper & Sample (Number of Galaxies) & Galaxy Mass $(M_\odot)$ & Redshift & Clump Finder & Detection Band \\ 
    \hline
    \hline
    E07 \citep{elmegreen07} & Starbursts (1003)  & N/A & $0<z<5$ & Visual & F775W \\  
    \hline
    P10 \citep{puech10} & Emission-line galaxies (63) & $>2 \times 10^{10}$ & $\sim 0.6$ & Visual & F435W \\  
    \hline
    O09 \citep{overzier09a} & Lyman Break Analogs (20) & $10^9 -- 10^{10}$ & $\sim 0.2$ & Visual & rest-frame UV \\  
    \hline
    G12 \citep{ycguo12clump} & Star-forming galaxies (10) & $>10^{10}$ & $1.5<z<2.5$ & Algorithm & F850LP \\  
    \hline
    W12 \citep{wuyts12} & Star-forming galaxies (649) & $>10^{10}$ & $0.5<z<2.5$ & Algorithm & rest-frame 2800\AA \\  
    \hline
    T14 \citep{tadaki14} & H$\alpha$-emitting galaxies (100) & $10^9 -- 10^{11.5}$ & $2.0<z<2.5$ & Algorithm & F606W \& F160W \\  
    \hline
    M14 \citep{murata14} & $I_{F814W}<22.5$ galaxies (24027) & $>10^{9.5}$ & $0.2<z<1.0$ & Algorithm & F814W \\  
    \hline
    G15 (Guo et al. in prep.) & Star-forming galaxies (50) & $>10^{10.75}$ & $0.05<z<0.25$ & Algorithm & F225W \\  
    \hline
    This work & Star-forming galaxies (3239) & $10^9 -- 10^{11.5}$ & $0.5<z<3.0$ & Algorithm & rest-frame 2500\AA \\  
    \hline
    \end{tabular}
\end{center}
\end{table*}

The new clumpy fraction ($f^{new}_{clumpy}$) depends on how many clumps ($n_c$)
a clumpy galaxy has. In the {\it bottom} panels of Figure \ref{fig:fclumpy}, we
plot the results with the assumption of $n_c=2$. Compared with the top panels,
although the amplitudes of $f_{clumpy}$ in different redshift and \mstar\ bins
are scaled up by, on average, a factor of $\sim$1.2, the trends with redshift
and \mstar\ are almost unchanged by taking into account the undetected clumps.
This is also true if we assume $n_c=1$, the most extreme case where each clumpy
galaxy only {\it intrinsically} has one clump. In that case, the amplitude will
be systematically scaled up by a factor $\sim$ 1.3, compared to the top panels. 

In this paper, we use \fclumpy\ under our default clump definition
($L_{blob}/L_{galaxy}=0.08$) and after the incompleteness correction with
$n_c=2$ as our best measurement (the bottom panels of Figure
\ref{fig:fclumpy}). Overall, low-mass galaxies ($\logm<9.8$) keep a constant
\fclumpy\ of $\sim$60\% from z$\sim$3 to z$\sim$0.5. Intermediate-mass galaxies
($9.8<\logm<10.6$) keep an almost constant \fclumpy\ of $\sim$55\% from
z$\sim$3 to z$\sim$1.5, and then gradually drops it to 40\% at $z\sim0.5$.
Massive galaxies ($10.6<\logm<11.4$) also keep their \fclumpy\ constant at
$\sim$55\% from z$\sim$3 to z$\sim$2, but then quickly drop it to $\sim$15\% at
z$\sim$0.5.

\subsection{Comparison with Other Studies}
\label{fclumpy:other}

We compare our $f_{clumpy}$ with that of other studies in the bottom right
panel of Figure \ref{fig:fclumpy}. The sample, \mstar\ range, and clump
identification method of each study used in the comparison are summarized in
Table \ref{tb:fclumpy}.

Our \fclumpy\ of $\logm>9.8$ galaxies shows good agreement with that of
\citet[][E07]{elmegreen07} and \citet[][P10]{puech10}, both identified clumpy
galaxies through visual inspection. E07 didn't specify the \mstar\ range of
their galaxies, but given their size and surface brightness cuts on the
rest-frame UV images of their galaxies, it is reasonable to compare their
results with our $\logm>10$ galaxies. Also, for E07, we only use their
categories of clump clusters, spirals, and ellipticals to calculate
$f_{clumpy}$. We exclude chain galaxies, double nuclei, and tadpoles, all of
which usually have small axial ratios, to match our requirement on the
elongation of galaxies. The agreement with the two measurements reinforces our
conclusion that $f_{clumpy}$ in massive galaxies drops from $\sim$50\% at
$z>1.5$ to about 20\% at $z\sim0.5$. 


The results of \citet[][M14]{murata14} also show agreement with our \fclumpy.
M14 measured \fclumpy\ for more than 20,000 galaxies at $0.2<z<1.0$ in COSMOS.
They identify clumpy galaxies through the peak of the contrast between the 1st,
2nd, and 3rd bright peaks in the F814W images of the galaxies. Their result at
$10.5<\logm<11.0$ galaxies (open triangles) shows good agreement with ours in
the highest \mstar\ bin (red circles). Their result at $10<\logm<10.5$ (open
diamonds) also matches ours in the intermediate \mstar\ bin (green diamonds) at
$z\sim$1. At $z<0.75$, however, their \fclumpy\ of $10<\logm<10.5$ galaxies
quickly drops, while we expect, from the extrapolation of our higher-redshift
results, a mild drop. More measurements are needed to confirm the trend of the
intermediate-mass galaxies at very low redshift.

Our \fclumpy\ is also statistically consistent with the results of two other
studies, given their large errobars. \citet[][T14]{tadaki14} measured \fclumpy\
for 100 H$\alpha$-emitting galaxies at $z=2.19$ and $z=2.53$. They identified
clumps from both F606W and F160W images, using the clump finder of
\citet{williams94} and adjusting the control parameters to match visual
inspections. Their sample spans a large \mstar\ range from $\logm=9.5$ to
$\logm=11.0$. Overall, their \fclumpy\ of 40\% is lower than ours, if we
combine all our \mstar bins together. \citet[][G12]{ycguo12clump} used an
algorithm similar to ours to identify clumpy galaxies in massive galaxies at
$z\sim2$. Their \fclumpy\ of 67\% is higher than ours, but their sample
contains only 15 galaxies, which is biased toward bright, blue, and large
galaxies because they only include spectroscopically observed galaxies.  Both
T14 and G12, however, have large uncertainties in their \fclumpy, making their
results still statistically consistent with ours.

\fclumpy\ of \citet[][W12]{wuyts12} are significantly higher than ours.
Instead of detecting individual clumps, W12 detect the pixels with excess
surface brightness in multi-band images. Here we use their \fclumpy\ measured
in the rest-frame UV detection. Their threshold of clumpy galaxies is quite low
compared to our definition here. They required a total UV (2800\AA) luminosity
contribution of 5\% from all clumps to be a clumpy galaxy, while we ask for at
least one clump contributing 8\% of the luminosity. As shown by the top left
panel of Figure \ref{fig:fclumpy}, a lower threshold would include a lot of
small star-forming regions, which could explain the high $f_{clumpy}$ of W12.  

\begin{figure*}[htbp]
\includegraphics[scale=0.30,angle=0,clip]{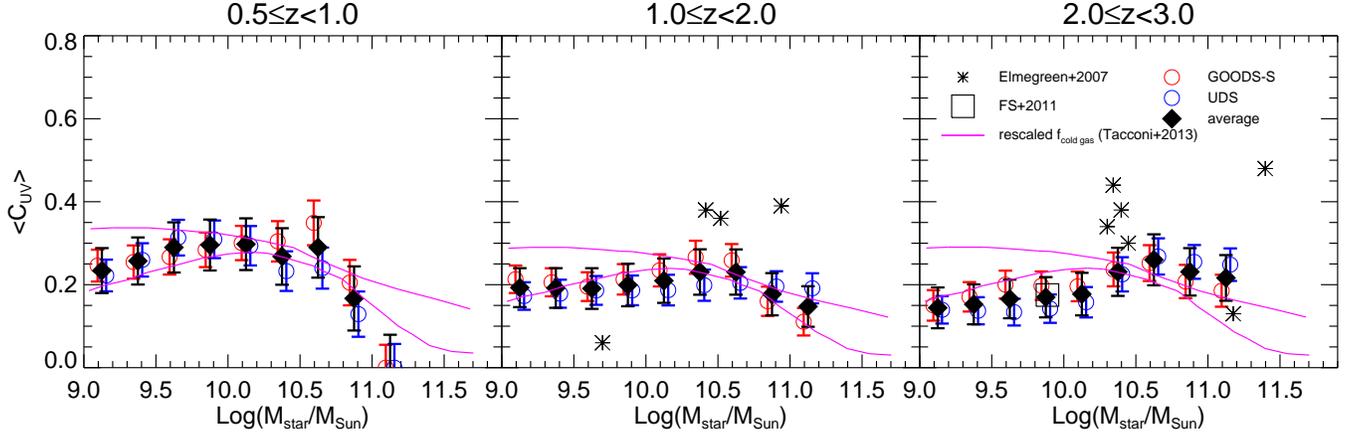}
\caption[]{Average clump contribution to the rest-frame UV light (\cuv, see
Sec. \ref{clumpfuv:vsms} for details) of a galaxy as a function of the \mstar\
of the galaxy. The contribution is corrected for the detection incompleteness
and averaged over all SFGs with or without detected clumps. The results of
GOODS-S (red circles and error bars), UDS (blue circles and error bars), and
the error-weighted average of the two fields (filled black diamond) are
plotted. Measurements from other studies are also shown, as the labels indicate
in the third panel. The ranges of the molecular gas fraction of
\citet{tacconi13} are normalized to match our fraction at $\logm=10.5$ and
overplotted with violet curves.
\label{fig:clfuvvsms}}
\vspace*{-0.2cm}
\end{figure*}

We also include the data of our ongoing \hst\ SNAPSHOT program
(HST-GO-13309) in Figure \ref{fig:fclumpy} as a boundary condition of the
clumpy fraction of massive SFGs at $z\sim0$. The SNAPSHOT program aims to image
a representative sample of 136 SDSS galaxies with $\logm>10.75$ and SSFR${\rm
>10^{-0.75}}$ at $0.05<z<0.25$ with the \hst/WFC3 UVIS F225W filter. The
details of the program and data reduction will be presented in a future paper
(Y. Guo et al., in preparation, G15). Here, we apply our blob finder to 50
galaxies that have been observed so far and identify clumps from them using the
same fractional luminosity threshold ($>$8\%). The clumpy fraction (the light
blue triangle with a circle embedded at $z\sim0.15$ in the lower right panel of
Figure \ref{fig:fclumpy}) confirms the rapid decline of \fclumpy\ in massive
galaxies -- only less than 10\% of massive SFGs at $z\sim0.15$ contain
off-center UV clumps. Among all other \fclumpy\ compared in Figure
\ref{fig:fclumpy}, this local sample has the closest sample selection,
observational effects, and clump identification to our CANDELS sample.
Therefore, it provides the most consistent constraint on the end-point of
massive \fclumpy\ evolution.

\subsection{Implications on Clump Formation}
\label{fclumpy:implication}

The fraction of clumpy galaxies and its evolution with redshift have important
implications for the formation mechanisms of the clumps. In a widely held view
based on theoretical works and numerical simulations, the clumps are formed
through gravitational instability in gas-rich turbulent disks
\citep[e.g.,][]{noguchi99,immeli04a,immeli04b,elmegreen08,dekel09,ceverino10,ceverino12,dekel14}.
This scenario is supported by the fact that high-redshift galaxies are
gas-rich, with gas--to--baryonic fraction of 20\% to 80\%
\citep[e.g.,][]{erb06,genzel08,tacconi08,tacconi10,fs09,daddi10}, possibly as a
result of smooth and continuous accretion of cold gas flow
\citep{keres05,rauch08,dekel09gas,cresci10,steidel10,giavalisco11}.
\citet{genzel11} derived the Toomre $Q$ parameter \citep{toomre64} at locations
of observed clumps from the H$\alpha$ velocity map and H$\alpha$ surface
density. They found $Q<1$, meaning gravitationally unstable to collapse, for
all clump sites and $Q\sim1$ throughout the disks, providing evidence for the
scenario of the violent disk instability \citep[VDI;][]{dekel09}.  The
kinematic signatures of the clumpy disks, however, can also have an ex-situ
origin, such as gas-rich mergers
\citep[e.g.,][]{robertson08,puech10,hopkins13}.

A few possible formation mechanisms of clumps are compared in Figure
\ref{fig:fclumpy_merger}. First, we compare the trend of disk instability
predicted by VDI with the evolution of $f_{clumpy}$ of massive galaxies
($\logm>10.6$), testing the idea that clumps are formed in-situ in turbulent
disks and are manifestations of gravitational instability in galaxy disks.
\citet{vanderwel14shape} found that about 80\% of the massive
($10.5<\logm<11.0$) SFG at $0<z<2$ are disky galaxies, which validates the
basic assumption of VDI, i.e., the existence of disks. Here, we use the
fiducial two-component (gas + stars) model of \citet{cacciato12} as
representative of VDI. In this model, the massive disk is assumed to be
continuously fed by cold gas at the average cosmological rate.  The gas forms
stars and is partly driven away by stellar feedback. The gravitational energy
released by the mass inflow down the gravitational potential gradient drives
the disk turbulence that maintains the disk instability. Since the gas is the
main driver of instability at all times, once the gas velocity dispersion
($\sigma_{gas}$) is significantly lower than the circular velocity of the disk
($V_{circ}$), the disk can be considered ``stable''. 

Since \citet{cacciato12} presents an analytic model with only one
realization for each set of parameters, we need to convert their model
prediction into a probability (i.e., some fraction of the galaxies) to make a
direct comparison with our \fclumpy. To this purpose, we assume that the trend
of $\sigma_{gas} /V_{circ}$ in \citet{cacciato12} is the average value for
massive SFGs. We then measure the scatter of the $\sigma_{gas} /V_{circ}$ from
the observations of \citet{kassin12}, who measured kinematics for a sample of
SFG at $z=0.2--1.2$.  The 1$\sigma$ scatter of $\sigma_{gas} /V_{circ}$ in
\citet{kassin12} is about 0.5 dex. A Monte Carlo sampling is then carried out
based on the average value and scatter to generate a distribution of
$\sigma_{gas} /V_{circ}$ at different redshifts. We then choose $\sigma_{gas}
/V_{circ} > 1/3$ as the threshold of being ``unstable disks'' (the same
threshold of \citet{kassin12}). We then compare the ``unstable'' fraction of
the Monte Carlo realizations with \fclumpy\ of our galaxies. The solid line in
Figure \ref{fig:fclumpy_merger} shows that the ``unstable'' fraction matches
\fclumpy\ of massive SFGs remarkably well, suggesting that VDI is a likely
explanation of the decrease of $f_{clumpy}$ toward low redshift. It is
important to note that, however, the VDI model used here does not directly
predict the threshold of forming clumps in an unstable disk. A more direct test
of VDI would be either comparing \fclumpy\ with the clump formation probability
predicted by VDI or comparing the characteristic \mstar\ of real clumps with
that of models (i.e., Toomre Mass predicted by VDI).

The trend of the disk stabilization in \citet{cacciato12}, however, cannot
explain the trend of the intermediate or low-mass galaxies. For the
intermediate ($9.8<\logm<10.6$) galaxies, the slow decrease of the trend at low
redshift requires the disk stabilization to be delayed. Although a few variants
of the fiducial model of \citet{cacciato12} are able to delay the disk
stabilization, e.g., by using an extremely high inflow accretion rate or an
extremely low star formation efficiency, their trend of disk instability
($\sigma_{gas} / V_{circ}$) cannot match the trend of the intermediate galaxies
at both low and high redshifts simultaneously. Other mechanisms may be needed
to explain the redshift evolution of $f_{clumpy}$ of the intermediate or
low-mass galaxies. It is important, however, to note that the models of
\citet{cacciato12} and similar VDI predictions are made for galaxies with
$\logm \sim 11$. Currently, no reliable predictions of VDI on the disk
stabilization for lower mass galaxies ($\logm<10.5$) are available. The general
VDI models \citep[e.g.,][]{bournaud11,dekel14}, however, do expect less massive
galaxies to remain unstable for longer times, because they retain higher gas
fractions as a result of the regulation of gas consumption in low-mass galaxies
\citep{dekel86,krumholz10} and the continuation of gas accretion to low
redshift for low-mass halos \citep{dekel06}.

In fact, \citet{elmegreen09b} made a qualitative analysis on the validity
of the gravitational instability in disks for local dwarf irregulars. They
found a striking resemblance of the morphology between high-redshift clumpy
galaxies and dwarf irregulars, although the former is intrinsically brighter
than the latter by a factor of 10--100. The typical velocity dispersion of
local dwarf irregulars is $~15 {\rm km s^{-1}}$, but their circular velocity is
small too, $< 100 {\rm km s^{-1}}$. As a result, the $\sigma_{gas}/V_{circ}$ of
dwarf irregulars is actually comparable to that of high-redshift disk galaxies
\citep{fs09}, indicating that they may be subject to the same physical
mechanism (VDI) to form disk. Future quantitative VDI models for low-redshift
low-mass galaxies would provide more detailed tests to the clump formation in
this regime.

Minor mergers, on the other hand, provide a viable explanation of $f_{clumpy}$
of the intermediate ($9.8<\logm<10.6$) galaxies at $z<1.5$.  \citet{lotz11}
measured the minor merger rate of $\logm>10$ galaxies at $z<1.5$:
$0.27\times(1+z)^{-0.1} Gyr^{-1}$. In order to compare the merger rate with
$f_{clumpy}$, we multiply the merger rate by an observability timescale of
1.5, 2, and 2.5 Gyrs (dashed black lines from bottom to top in Figure
\ref{fig:fclumpy_merger}). The comparison shows that if the observability timescale of minor mergers is between 1.5 and 2 Gyr, the minor merger fraction
qualitatively matches $f_{clumpy}$ of $9.8<\logm<10.6$ galaxies.  In fact,
\citet{lotz10b} studied the effect of gas fraction on the morphology and
timescales of disk galaxy mergers and found that when the gas fraction is
higher than 50\%, the timescales for morphological disturbances measured by
Gini and/or Asymmetry can be as long as or more than 1.5 Gyrs for a 9:1
baryonic mass ratio merger. Given the high gas-fraction of galaxies around
$\logm=10$ at $z\sim1$ \citep[e.g.,][]{tacconi13}, the long observability timescale of minor mergers is feasible.  Therefore, minor mergers are a viable
explanation of $f_{clumpy}$ of the intermediate ($9.8<\logm<10.6$) galaxies at
$z<1.5$. At $z>1.5$ as well as for low-mass ($\logm<9.6$) galaxies at all
redshifts, since the minor merger rate has not been studied thoroughly
\citep[e.g.,][]{newman12}, it is hard to evaluate the role of minor mergers on
clump formation at $z>1.5$.

The connection between minor mergers and clumps can happen in two ways.  First,
minor mergers are the clumps themselves. Such clumps are called ex-situ clumps,
while those formed in the disk are called in-situ clumps.  \citet{mandelker14}
analyzed the cosmological hydro-simulations of \citet{ceverino10} and found
that about 15\% of clumps are ex-situ clumps (i.e., they have their own dark
matter components before merging into the primary galaxies). They also found
that ex-situ clumps have higher mass, lower gas fraction, lower specific SFR,
and older stellar population than in-situ clumps have. We will carry out a
detailed analysis on the properties of clumps in a future paper to test the
fraction of ex-situ clumps in observations.  Second, minor mergers can induce
clump formation in the disks.  Minor merger would disturb the cold gas
distribution in the disks and increase the gas surface density locally, which
then results in a local mini-starburst to form a clump (see the resemblance of
clumps and starbursts in \citet{bournaud14}).  Unlike gas-rich major mergers,
which induce galaxy-wide star-bursts, each minor merger may only be able to
induce local starbursts in a few locations. S. Inoue et al. (in preparation)
found that in simulations, clumps can be formed in some stable (Toomre Q$>$1)
disks. In this case, external stimulation (e.g., minor merger) may be needed to
enhance the gravitational instabilities to form clumps.

Another possible clump formation mechanism --- major mergers --- is unlikely
able to explain $f_{clumpy}$ evolution in any mass range, unless its
observability timescale is $\gtrsim$3 Gyrs. Figure \ref{fig:fclumpy_merger}
shows that only with such a long timescale, is the major merger rate from
\citet{lotz11} ($0.03\times(1+z)^{1.7} Gyr^{-1}$) able to match $f_{clumpy}$ of
$\logm>10.6$ galaxies at $z<1.5$. \citet{lotz10b}, however, shows that the
timescales for morphological disturbances measured by various merger indicators
are all less than 2 Gyr even when the gas fraction of the equal-mass merger is
as high as 60\%. Longer timescales require even higher gas fractions.  At
$z\sim1$, however, \citet{tacconi13} show that the gas fraction of massive
galaxies is $\lesssim$30\%, which implies a much shorter ($\sim$0.5 Gyr)
observability timescale for massive galaxies. Another measurement of the wet
major merger fraction from \citet{ls13} is also lower than $f_{clumpy}$ at
$0<z<1.8$. Therefore, we conclude that major mergers are unlikely to be a
viable explanation for the observed trend of \fclumpy. 

In summary, VDI predicts a strongly decreasing trend of disk instability
towards the present day, which qualitatively matches the evolution of \fclumpy\
of massive galaxies. 
The normalization and the
slopes of the minor and major merger fractions are uncertain, depending on
their observability timescale, especially for the low mass galaxy bin. The
comparisons show some level of correspondence with the expectations of a minor
merger origin of clumps for intermediate-mass galaxies at $z<1.5$, but appear
to strongly disfavor a major merger origin of clumps (observability timescales
would need to exceed 3 Gyr) for all galaxies at $z<1.5$. The effects of both
minor and major mergers on the clump formation at $z>1.5$ are still unclear due
to the lack of a robust estimate of the merger rates. The expectations of all
scenarios would benefit from further more realistic modeling of larger
cosmologically-motivated samples of galaxies.

\section{Clump Contribution to the Rest-frame UV Light of Galaxies}
\label{clumpfuv}

\subsection{Clump Contribution to the Rest-frame UV Light of Galaxies}
\label{clumpfuv:vsms}

The clump contribution to the rest-frame UV light per galaxy ($C_{UV}$) can be
derived through integrating the FLF: $C_{UV} = \int n({\mathscr L}){\mathscr L} d{\mathscr L}$, where ${\mathscr L}=L_{clump}/L_{galaxy}$ and the upper and lower limits of the integration are
unity and the threshold of our clump definition (Sec. \ref{definition}),
respectively. Because our incompleteness-corrected FLFs take into account both
``clumpy'' and ``non-clumpy'' galaxies (see Sec. \ref{fuv} for the
discussion of why we cannot separate them), the $C_{UV}$ we discuss later is
the contribution to the entire population of SFGs averaged over both ``clumpy''
and ``non-clumpy'' galaxies. 

One issue has to be considered when we measure $C_{UV}$: the blending
of clumps and blobs in high-redshift and/or low-mass bins, where galaxies tend
to have smaller sizes (Figure \ref{fig:sample}). The blob blending
gradually shifts the FLF to the bright end, but meanwhile lowers the peak of
the FLF. The net result of the two actions is that the {\it number of the
clumps} are kept more-or-less the same. The clump luminosity is, however,
artificially increased due to the blending, which would result in an
overestimate of \cuv\ if we simply integrate the observed FLF. \footnote{This 
issue has little effect on the measurement of $f_{clumpy}$. We confirm
this in the top right panel of Figure \ref{fig:fclumpy}. In this panel, we
measure a new \fclumpy\ by replacing \cuv\ in Equation \ref{eq:flfratio} and
\ref{eq:flfmeasure} by \fclumpy. The new results (dashed lines) show that the
normalization and trend of galaxies at $0.5<z<2$ are almost preserved. Only for
massive ($\logm > 10.6$) galaxies at $2<z<3$, the new \fclumpy\ apparently
deviates from the old \fclumpy, suggesting an almost flat \fclumpy--\mstar
relation at $z>2$ and an earlier decline of \fclumpy\ for massive galaxies
starting from $z\sim3$ in the \fclumpy\--redshift trend.}

To solve this issue, we use a different way, instead of integration of the
incompleteness-corrected FLF, to calculate $C_{UV}$. We choose the galaxies
with $0.5<z<1.0$ and $9.8<\logm<10.6$ as fiducial templates, assuming (1) the
clump--clump or clump--blob blending issue is the least significant in this
(\mstar, $z$) bin and (2) galaxies and clumps in this (\mstar, $z$) bin are
representative for all galaxies and clumps in our sample.  We then rescale the
size and apparent magnitude of the template galaxies to match the median size
and magnitude (in units of ADU) of galaxies in other (\mstar, $z$) bins. We
then run our blob finder on the rescaled galaxies and derive the FLF in each
bin.  The newly derived FLFs, as shown by blue histograms in Figure
\ref{fig:compm101}, tell us the clump contribution if we move the fiducial
galaxies to other (\mstar, $z$) bins, taking into account both the size change
as well as the cosmological dimming.

For each (\mstar, $z$) bin, its actual clump contribution, $C_{UV} (M_{*},z)$,
is then derived by scaling the clump contribution of the fiducial bin, $C_{UV}
(M_{*F},z_{F})$, by a ratio $\Upsilon$:
\begin{eqnarray} 
C_{UV} (M_{*},z) = \Upsilon C_{UV} (M_{*F},z_{F}).  
\label{eq:flfmeasure} 
\end{eqnarray}  

\noindent $\Upsilon$ is defined as 
\begin{eqnarray}
\Upsilon 
& \equiv & \frac{c^{\prime}_{UV}(M_{*},z)}{c^{\prime,rs}_{UV}(M_{*F},z_{F})} 
= \frac{\it X c^{\prime}_{UV}(M_{*},z)}{\it X c^{\prime,rs}_{UV}(M_{*F},z_{F})} \nonumber \\ 
& = & \frac{C_{UV}(M_{*},z)}{C^{rs}_{UV}(M_{*F},z_{F})},
\label{eq:flfratio}
\end{eqnarray} 

\noindent where $c^{\prime}_{UV}(M_{*},z)$ and
$c^{\prime,rs}_{UV}(M_{*F},z_{F})$ are the clump contributions of the FLF in
the (\mstar, $z$) bin and of the redshifted fiducial FLF {\it before} the
detection incompleteness is corrected. We use the ratio of
$c^{\prime}_{UV}(M_{*},z)$ and $c^{\prime,rs}_{UV}(M_{*F},z_{F})$ to measure
$\Upsilon$. The incompleteness correction factor in the (\mstar, $z$) bin ($\it
X$ in the above equation) is the same for both the actual and the redshifted
fiducial FLFs, because both have the same matched observational effects and
galaxy surface brightness distributions. Therefore, the measured $\Upsilon$ is
equal to the ratio of $C_{UV} (M_{*},z)$ and $C^{rs}_{UV} (M_{*F},z_{F})$,
i.e., the ratio of the clump contributions of the FLF in the (\mstar, $z$) bin
and of the redshifted fiducial FLF {\it after} the detection incompleteness is
corrected. If $\Upsilon$ is larger than one, the clump contribution in the
(\mstar, $z$) bin is higher than that in the fiducial bin, and vice versa.
Since the intrinsic clump contribution in the fiducial bin would not change
when the galaxies are redshifted, we have $C_{UV} (M_{*F},z_{F}) = C^{rs}_{UV}
(M_{*F},z_{F})$, which leads Eq. \ref{eq:flfratio} back to Eq.
\ref{eq:flfmeasure}.

It is important to note that Figure \ref{fig:compm101} is only for illustrating
the comparison between actual and redshifted FLFs. In reality, we use finer
(\mstar, $z$) bins to derive $C_{UV} (M_{*},z)$. Also, we do not apply this
method to galaxies with $\logm>10.6$ at $z<1$, because their sizes are larger
than that of the fiducial galaxies. The blending issue is even less a problem
for them than for the fiducial galaxies.

An example of how this method overcomes the clump--clump or clump--blob
blending issue can be seen from the $2.0<z<3.0$ and $9.0<\logm<9.8$ bin in
Figure \ref{fig:compm101}. If we simply integrate the incompleteness-corrected
FLFs (black histograms) of both this bin and the fiducial bin down to the clump
definitions, the clump contribution of this bin would be larger than that of
the fiducial bin. The blue histogram in the figure, however, shows that once we
redshifted the fiducial galaxies to match the size and magnitude of the
galaxies in this (\mstar, $z$) bin, their FLF is actually significantly higher
than that of the real galaxies in this bin in the bright end, implying that the
clump contribution of real galaxies in this bin is actually lower than that in
the fiducial bin. The reason for the apparently higher clump contribution of
this high-redshift bin is actually due to the clump--clump or clump--blob
blending, which makes clumps look large/bright.

Figure \ref{fig:clfuvvsms} shows $C_{UV}$ as a function of \mstar\ in three
redshift bins. $C_{UV}$ does not change monotonically with \mstar. It increases
with the \mstar\ from $\logm \sim9.0$ to $\logm \sim10.0$, and then reaches a
broad peak around $\logm \sim10.5$, and quickly drops at $\logm >10.8$. 


We compare our measurements with other studies in Figure
\ref{fig:clfuvvsms}. \citet{elmegreen05} measured the light contribution of
ten galaxies at $1<z<3$ in the $i$-band (F775W) in the HUDF. \citet{fs11b} also
measured the F775W light contribution of one galaxy in their $z\gtrsim2$
sample. Both studies counted only ``clumpy'' galaxies. It is then not
surprising that their clump contributions are higher than ours.  At $1<z<3$,
the values of \citet{elmegreen05} are indeed systematically higher than the
average of our two fields. The median of their work is about 2 times higher
than our value in the $\logm \sim10.5$ mass bin. The value of the single galaxy
of FS11, in the $\logm \sim9.9$ mass bin, is slightly higher than our
measurements in both fields.  If we assume that all our clump contribution is
from the detected ``clumpy'' galaxies and use $f_{clumpy} = 0.5$ from Figure
\ref{fig:fclumpy_merger}, our clump contribution at $10<\logm<11$ and $1<z<3$
should be scaled by a factor of 2, when we only consider the clump
contributions to clumpy galaxies. Our $C_{UV}$ is then consistent with that of
the two studies. Overall, our measurements are broadly consistent with those of
other studies, but the small samples of other studies prevent us from making
robust comparison at all redshift and \mstar\ ranges.

\subsection{A Possible Link to Molecular Gas Fraction}
\label{clumpfuv:fgas}

The trend of $C_{UV}$ as a function of the galaxy \mstar\ could be mostly
driven by the molecular gas fraction of the galaxies. The physical link could
be established in two ways. First, UV bright clumps are shown to be regions
with enhanced specific star-formation rates \citep{fs11b, ycguo12clump,
wuyts12, wuyts13}. Since the star formation activity is controlled by the
amount of molecular gas in galaxies, it is natural to assume that $C_{UV}$, a
signal of the strength of star formation in galaxies, reflects the molecular
gas fraction in these galaxies. Second, an important condition of forming
clumps is gas-rich galaxies in both the VDI and merger scenarios. Therefore,
how important the clumps are in the galaxies, namely the fraction of the stars
that formed in the giant clumps, is determined by the cold gas fraction of the
galaxies\footnote{Here we use molecular gas fraction to represent the cold
gas fraction. This is valid if the molecular gas mass dominates the cold gas
mass, which is true for galaxies at z$>$1.5 \citep{obreschkow09} as well as for
massive galaxies ($>$a few times $10^{10} M_\odot$) at $z\sim1$
\citep{tacconi13}.}.

To test the possible link between the gas fraction and $C_{UV}$, we
overplot the molecular gas fraction of \citet{tacconi13} in Figure
\ref{fig:clfuvvsms}. \citet{tacconi13} measured the molecular gas fraction for
50 star-forming galaxies at z$\sim$1--1.5. Here we use their result that takes
into account the incompleteness of the PHIBSS survey. The gas fraction of
galaxies below their detection limit was derived by using an empirical gas
depletion timescale. \citet{tacconi13} also showed that once the gas fraction
is normalized to the value at $M_{*}=10^{10.5} M_\odot$, the gas
fraction--\mstar\ trend of z=0 SFGs and that of z$\sim$1--1.5 SFGs show
remarkable agreement with the redshift dependence almost fully removed.
Therefore, assuming that the gas fraction is responsible for the clump
contribution $C_{UV}({\rm M_{*}}, z)$, we simply rescale the gas fraction at
$z=1$ in order to match $C_{UV}$ in all redshift bins: 
\begin{eqnarray}
C_{UV}({\rm M_{*}}, z) & = & A \times f_{gas}({\rm M_{*}}, z) \\
                  & = & a(z) \times \frac{f_{gas} ({\rm M_{*}}, z=1)}{f_{gas} (10^{10.5}M_\odot, z=1)},
\label{eq:fgas}
\end{eqnarray}
where $a(z)$, the normalization factor, is redshift dependent and may be
determined by other physical mechanisms. 

The normalized molecular gas fraction shows agreement with our $C_{UV}$ in the
intermediate and massive \mstar\ bins, both showing a quick drop above $\logm
=10.6$. This supports our speculation that $C_{UV}$ is mostly driven by the
molecular gas fraction of galaxies, at least for intermediate and massive
galaxies at $0.5<z<2$. Since There are almost no robust measurements of
the gas fraction--\mstar\ relation at $2<z<3$, it is unclear if our comparison
with the normalized lower-redshift molecular gas fraction is still valid at
$2<z<3$.

The link between \cuv\ and the molecular gas fraction for massive galaxies
at $0.5<z<2$ is also consistent with our previous result that VDI is likely
responsible for the quick drop of \fclumpy\ of massive galaxies toward low
redshift. VDI predicts that the lower the cold gas surface density, the more
stable the disk. For massive galaxies with a given \mstar\ (e.g.,
$10^{11}M_\odot$), toward low redshift, the cold gas fraction decreases
\citep{saintonge11} but the size of galaxies slightly increases
\citep{vanderwel14size}. Therefore, the cold gas surface density decreases and
hence results in a stable disk where clumps are hard to form. In fact, the
model of \citet{cacciato12} used in Fig. 11 takes into account of the gas
surface density decreases. In their model, the Toomre $Q$ parameter at low
redshift is dominated by the stellar component rather than by the gas
component.

It is difficult to draw firm conclusion on whether the same
\cuv--molecular gas fraction link is still valid for low-mass ($<10^{10}
M_\odot$) galaxies due to a few reasons. First, when measuring \cuv, we assume
that (1) clumps in galaxies at different (\mstar, $z$) bins are self-similar
and (2) the clump FLF in the fiducial bin is representative for the intrinsic
FLFS of all other bins. The two assumptions may break down preferentially for
low-mass galaxies because of their small sizes (Figure \ref{fig:sample}).
Second, there is no robust measurement of the molecular gas fraction in
low-mass galaxies. \citet{tacconi13} only measured molecular gas fraction for
massive galaxies. Their fraction for galaxies with $\logm<10$ is calculated
through $f_{gas} = 1/(1+1/(SSFR\times t_{depl}))$, where $SSFR$ is the average
specific SFR of the star-forming main sequence in literature and $t_{depl}$ is
the gas depletion timescale. \citet{saintonge11} found that $t_{depl}$ is
positively correlated with \mstar, but \citet{tacconi13} set a constant
$t_{depl} = 0.7 Gyr$ for all galaxy masses. Therefore, low-mass galaxies should
have lower $t_{depl}$ and hence lower inferred gas fraction than in
\citet{tacconi13}. A further measurement of molecular gas fraction for low-mass
galaxies is needed to investigate the \cuv--molecular gas fraction in the
low-mass end, which can be used to study the physical mechanisms of regulating
star formation \citep[e.g., radiation pressure feedback discussed
by][]{moody14}.

\begin{figure}[htbp]
\hspace*{-0.4cm}
\includegraphics[scale=0.45,angle=0,clip]{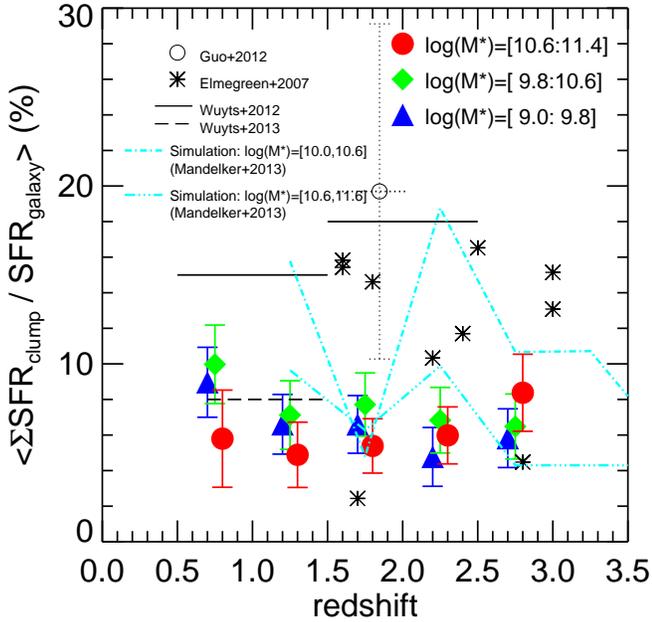}
\caption[]{Clump contribution to the SFR of galaxies. The colored points with
error bars are from our results, divided into different galaxy \mstar\ bins, as
the labels show. Measurements from other studies are also overplotted.  Note
that our study and \citet{wuyts12,wuyts13} measured the clump contribution to
the SFR of all galaxies, both clumpy and non-clumpy, while \citet{ycguo12clump}
and \citet{elmegreen05} measured the clump contribution to the SFR of only
clumpy galaxies. The results of \citet{mandelker14}, who analyzed the
cosmological hydro-simulation of \citet{ceverino10}, are also plotted, after
being converted to the contribution to all galaxies.
\label{fig:clsfr}}
\end{figure}

\subsection{Clump Contribution to the Cosmic Star Formation}
\label{clumpfuv:sfr}

We derive the clump contribution to the SFR of SFGs (\csfr) from \cuv. The
rest-frame UV light is a good tracer of SFR if the dust extinction is well
understood. In our case, we need to know the dust extinction of both clumps and
intra-clump regions. Such dust extinction maps can be measured through
spatially resolved multi-band SEDs
\citep[e.g.,][]{ycguo12clump,wuyts12,wuyts13}. We leave accurate measurements
of clump properties to a future paper. Instead, we try to use a simplified way
to derive a first order estimate of \csfr. 

We assume that the dust extinction difference between clumps and galaxies is
$\Delta A_V = A_V(galaxy) - A_V(clump) = 0.6$ and scale down \cuv\ by a factor
of $f_{dust} = 10.0^{-0.4 \times \Delta A_V/R_V \times k(2500\AA)} = 0.34$,
where $R_V$ and $k(2500\AA)$ are derived from the Calzetti Law
\citep{calzetti00}.  This prescription is predicated on two assumptions.
First, since we only use galaxies with axial ratio $q>0.5$, we assume the dust
extinction in our sample is quite small, with a median value of $A_V=0.6$.
Second, as shown by \citet{wuyts13}, UV bright clumps are regions with smaller
dust extinction than the intra-clump regions. 
Based on our previous study \citep{ycguo12clump}, we believe that the
extinction difference of $\Delta A_V = 0.6$ is valid for most of the clumps and
galaxies. On the other hand, if we assume that the dust extinction of the clumps
and the overall galaxies are the same, i.e., $\Delta A_V = A_V(galaxy) -
A_V(clump) = 0.0$, \csfr\ would be equal to \cuv, namely, $f_{dust}=1$. In this
case, all values of \csfr\ in our later discussions should be scaled up by a
factor of $\sim$3.

The redshift evolution of \csfr\ for different mass bins is shown in Figure
\ref{fig:clsfr}.  For low- and intermediate-mass galaxies ($9<\logm<10.6$),
\csfr\ increases from 6\% at $z\sim3$ to 10\% at $z\sim0.5$. For massive
galaxies ($\logm>10.8$), \csfr\ shows a mild decrease from 8\% at $z\sim3$ to
5\% at $z\sim0.5$.

It is interesting to compare our results with other studies. Both
\citet{ycguo12clump} and \citet{elmegreen05} studied clumps in massive galaxies
($\logm>10$) at z$>$1.5. Since their samples only contain clumpy galaxies, their
\csfr\ should be higher than our value.  \citet{ycguo12clump} measured the SFR
of each clump through spatially resolved SED-fitting. We include the
contribution of all off-center clumps from their sample. Their \csfr\ (circle
in Figure \ref{fig:clfuv_fit}) is about 20\%, a factor of 2.5 higher than our
values of massive ($\logm>10$) galaxies at $z\sim2$. \citet{elmegreen05} only
measured the F775W light fraction from clumps.  We adopt the above assumption
$\Delta A_V = A_V(galaxy) - A_V(clump) \sim A_V$ to convert their UV light
fraction into SFR fraction. Their values (stars in the figure) are about two
times higher than ours. The higher \csfr\ of both papers is broadly consistent
with our expectation, because they only consider the contribution to ``clumpy''
galaxies, while we consider the contribution to all SFGs.

\citet{wuyts12} measured \csfr\ of $\logm>10$ galaxies at $0.5<z<2.5$ in the
GOODS-S field through spatially resolved SED-fittings. Because their sample
contains both clumpy and non-clumpy galaxies, their \csfr\ provides a direct
comparison to our study.  Here we only quote their measurements based on the
rest-frame 2800\AA\ clump detection.  Their values are higher than ours by a
factor of $\sim$1.5.  It is important to note that \citet{wuyts12} did not
detect each individual clump. Instead, they focused on regions with excess
surface brightness and only statistically separated clump pixels from disk and
bulge pixels. Also, they did not subtract background when measuring clump
fluxes. Both could contribute to the discrepancy.  \citet{wuyts13} revisited
the clump contribution to SFR by combining CANDELS and 3D-HST
\citep{brammer123dhst}.  They derived the SFR of clump and disk pixels from
dust-corrected H$\alpha$ luminosity. At $0.7<z<1.5$, their \csfr\ of $\logm>10$
galaxies drops from 15\% in \citet{wuyts12} to 9\%, and is now in very good
agreement with ours.


\subsection{Comparison with Cosmological Hydrodynamic Simulations}
\label{fclumpy:sim}


We compare our \csfr\ with that of the state-of-art numerical simulations.
\citet{mandelker14} analyzed a large sample of simulations, generated by the
same code of \citet{ceverino10}, to detect clumps from snapshots of 3D gas
density. In Figure \ref{fig:clsfr}, we overplot (in cyan) their \csfr\ for two
\mstar\ bins. The curves are made by scaling down \csfr\ of ``clumpy'' galaxies
in \citet{mandelker14} by \fclumpy\ in their simulations.  Their \csfr\ shows
no clear trend from z=3.5 to z=1.0.  The large fluctuations of the curves,
which reflect the uncertainty levels of their measurements, prevent us from
drawing firm conclusions. Overall, their \csfr\ of intermediate-mass
($10.0<\logm<10.6$) galaxies seems higher than our measurements, while their
\csfr\ of massive galaxies ($10.6<\logm<11.6$) agrees with ours within the
uncertainties. 

But the definitions of clumps in our paper and \citet{mandelker14} are
different. Besides identifying clumps from 3D gas snapshots,
\citet{mandelker14} included lots of small clumps with SFR contribution less
than a few percent of that of the galaxies. The different clump definitions
could make the above comparison unfair. A proper way to compare observations
with simulations is to generate simulated images that match all the
observational effects of the real images. An example can be found in
\citet{moody14}. The same set of simulations has been run through SUNRISE
\citep{jonsson06,jonsson10a,jonsson10b} to calculate the radiative transfer and
then generate light images in given observational bands. These light images are
then downgraded to match the resolution and noise level of real CANDELS images
(called ``CANDELization'', see \citet{snyder14} and M. Mozena et al. in preparation). In a separate
paper, we will run our clump finder on the ``CANDELized'' simulation images to
make a direct comparison of clumpy galaxies between observations and
simulations.

\section{Conclusions and Discussions}
\label{conclusion}

In this paper, we measure the fraction of clumpy galaxies in SFGs and the clump
contributions to the rest-frame UV light and SFR of SFGs in the CANDELS/GOODS-S
and UDS fields. Our mass-complete sample consists of 3239 SFGs (SSFR$>$0.1
Gyr$\rm ^{-1}$) at $0.5<z<3$ with axial ratio $q>0.5$.  We propose a definition
of the UV-bright clumps in a way that is more physical than the appearance of
galaxies and is easier to apply to other observations and model predictions.
Our main conclusions are summarized below:

\begin{enumerate}

\item We define clumps as discrete star-forming regions that individually
contribute more than 8\% of the rest-frame UV light of their galaxies. This
definition is determined by comparing the fractional luminosity function of
star-forming regions, i.e., the number of star-forming regions per galaxy that
contribute a given fraction of the UV luminosity of the galaxies, of real and
redshifted nearby spiral galaxies. Clumps defined this way are significantly
brighter than the redshifted HII regions of nearby large spiral galaxies
and hence cannot be explained by the blending of the HII regions due to the
decrease of physical spatial resolution and cosmological dimming.

\item Given the above definition, we measure the fraction of clumpy galaxies
(\fclumpy) in SFGs, requiring each clumpy galaxy to contain at least one
off-center clump. The redshift evolution of the clumpy fraction changes with
the \mstar\ of the galaxies. Low-mass galaxies ($\logm<9.8$) keep a constant
\fclumpy\ of $\sim$60\% from z$\sim$3 to z$\sim$0.5. Intermediate-mass galaxies
($9.8<\logm<10.6$) keep their \fclumpy\ almost a constant around 55\% from
z$\sim$3 to z$\sim$1.5, and then gradually drops it to 40\% at $z\sim0.5$.
Massive galaxies ($10.6<\logm<11.4$) also keep their \fclumpy\ constant at
$\sim$55\% from z$\sim$3 to z$\sim$2, but then quickly drop it to $\sim$15\% at
z$\sim$0.5. 

\item \fclumpy\ decreases with \mstar\ at all redshift ranges, but the slope of
the decrease changes with the redshift: the lower the redshift is, the faster 
the trend decreases. At $0.5<z<1.0$, \fclumpy\ decreases from 60\% at $\logm \sim9.0$ to
15\% at $\logm \sim11.5$.  At $1<z<2$, \fclumpy\ decreases from around 60\% at
the lowest \mstar\ to $\sim$30\% at $\logm > 11.00$. At the highest redshift bin
$z>2$, \fclumpy\ only shows a mild decrease from $\sim$55\% at $\logm \sim9$
to 45\% at $\logm \sim10.5$. 

\item \fclumpy\ has important implications for the formation mechanisms of the
clumps. We find that (1) the trend of disk stabilization predicted by VDI
matches the \fclumpy\ trend of massive galaxies; 
(2) minor mergers are a viable explanation of the
\fclumpy\ trend of intermediate galaxies at $z<1.5$, given a realistic
observability timescale; and (3) major mergers are unlikely responsible for
\fclumpy\ in all masses at $z<1.5$. The roles of both minor and major mergers
on low-mass galaxies at all redshifts or on intermediate-mass and massive
galaxies at $z>1.5$ are still unclear due to the lack of a robust estimate of
the merger rates at $z>1.5$. 

\item We derive the clump contribution to the total UV luminosity of the
galaxies (\cuv), taking into account the effects of clump--clump and
clump--blob blending at high redshifts. At all redshifts, \cuv\ increases with
the \mstar\ of the galaxies from $\logm \sim 9$ to $\logm \sim 10$, reaches to a broad peak around $\logm \sim10.5$,
and then quickly drops. We speculate that the molecular gas fraction plays a
major role on the trend of \cuv\ in intermediate and massive galaxies at least.

\item We convert \cuv\ into the clump contribution to the SFR of the SFGs
(\csfr), under an assumption that the dust extinction of clumps is lower than
that of the galaxies. The redshift evolution of \csfr\ shows mild trends at different
\mstar\ ranges. For low- and intermediate-mass galaxies ($9<\logm<10.6$),
\csfr\ increases by almost a factor of 2 from $z\sim3$ to $z\sim0.5$.
For massive galaxies ($\logm>10.8$), \csfr\ shows a mild decrease from $z\sim3$
to $z\sim0.5$. We emphasize again that both our \cuv\ and \csfr\ are the
contributions to all SFGs rather than to only ``clumpy'' galaxies.




\end{enumerate}

It is important to note that our clump definition is established from
observations and does not incorporate any prior requirements from any
theoretical models. For example, we do not ask our clumps to be gravitationally
bound or formed in disks. 
The detachment of the observations and theories is important when we are still
carrying out basic demographic studies of observational phenomena. It
ensures that the demographic results would not be biased by any prevalent
theoretical models, especially when the demographic results will be used to
test the validity of those models.

It should also be noted that all our conclusions are based on {\it off-center}
clumps. We do not include any clumps within $0.5 \times r_e$ of galaxies, which
may bias our results. For example, the low \fclumpy\ and low \cuv\ in
high-redshift low-mass galaxies could be due to the neglect of clumps with
small galactocentric distances, if clumps are preferentially formed or evolved
into locations close to the galactic centers. 

Our results are also only valid for {\it UV-bright} clumps. This is reasonable
given the fact that clumps are traditionally and mostly identified in the
rest-frame UV images \citep[e.g.,][]{elmegreen05,elmegreen07,ycguo12clump}.
Recent studies, however, also identify clumps from the rest-frame optical
images \citep[e.g.,][]{elmegreen09a,fs11b}. These ``red'' clumps contain
important clues to the fate of the clumps, namely, whether they would migrate
toward the gravitational centers of their host galaxies, due to clump
interactions and dynamical friction,  and eventually coalesce into a young
bulge as the progenitor of today's bulges
\citep[e.g.,][]{elmegreen08,ceverino10} or be quickly disrupted by either tidal
force or stellar feedback to become part of a thick disk
\citep[e.g.,][]{bournaud09,dekel09,murray10,genel12}. To answer this question,
a similar physical definition of clumps needs to be migrated to the red bands,
and an accurate measurement of clump properties (e.g., \mstar\, age, SFR) is
required.

 
\begin{figure*}[htbp]
\center{\includegraphics[scale=0.4, angle=0]{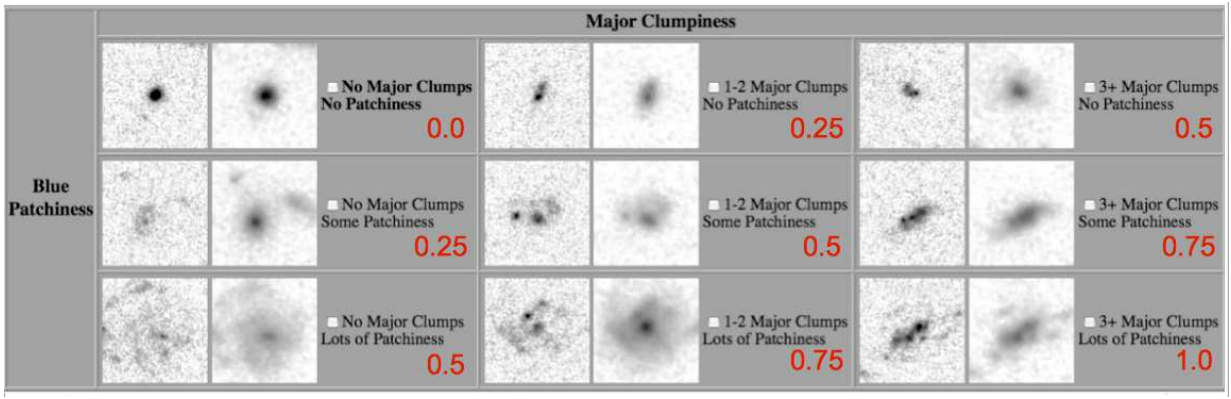}}
\caption[]{Scheme of the CANDELS visual classification of galaxy clumpiness.
For each galaxy, each inspector can choose more than one option in the
3$\times$3 matrix spanned by blue patchiness and major clumpiness. The score
of each option (cell) is labeled by the red numbers. The scores of the selected
options are then averaged to get the score of the inspector.
The scores of all inspectors are then averaged to get a single score between 0
and 1 for each galaxy.
For each option, the images are the F606W (left) and F160W (right) bands.
\label{fig:visual}}
\vspace{0.2cm}
\end{figure*}

\ \ \ 

\ \ \

We thank the anonymous referee for constructive comments that improve this
article. Support for Program number HST-GO-12060 and HST-GO-13309 were
provided by NASA through a grant from the Space Telescope Science Institute,
which is operated by the Association of Universities for Research in Astronomy,
Incorporated, under NASA contract NAS5-26555. We acknowledge partial support from NSF grant AST 08-08133.

{\it Facilities}: \hst\ (ACS and WFC3)

\vspace*{0.3cm}

\appendix

\twocolumngrid

\section{A. Comparison between Our Blob Finder and CANDELS Visual Clumpiness}
\label{clfinder:visual}

A sanity check of the efficiency and accuracy of our blob finder is to compare
it with visual inspections. The CANDELS team has carried out a visual
classification of galaxies with $H_{F160W} <$24.5 AB mag in the both GOODS-S
and UDS fields to determine the morphological class (i.e., spheroid, disk,
irregular, etc.), interaction class (i.e., merger, interacting, non-merger,
etc.), clumpiness, etc. of the galaxies. Each galaxy has been inspected by 3--5
astronomers, mainly through its F160W images and complemented by its images in
other bands. The details of the visual classification and the results are given
by \citet{kartaltepe14}. Here, we only use its visual clumpiness. 

\begin{figure*}[htbp]
\includegraphics[scale=0.3,angle=0, clip]{./fig15a.ps}
\includegraphics[scale=0.3,angle=0, clip]{./fig15b.ps}
\includegraphics[scale=0.3,angle=0, clip]{./fig15c.ps}
\caption[]{Comparison of the CANDELS visual clumpiness (left), visual patchiness (middle), and visual clumpiness plus patchiness (right) with the number of blobs
(bulge or galaxy center is excluded) detected by the automated blob finder. The
gray scale shows the number of galaxies (as indicated in the gray-scale bar in
the right side) in each bin. Red diamonds and blue circles show the mean and
median visual clumpiness at given number of blobs. Green curves from bottom to
top show the 10th, 20th, 80th, and 90th percentiles. 
\label{fig:vvsc}}
\end{figure*}

The scheme of visually determining clumpiness is shown in Figure
\ref{fig:visual}. It starts from a 3$\times$3 grid spanned by blue patchiness
and major clumpiness. Each cell of the grid is given a score as the following:
\begin{enumerate} 
\item[0.0:] no clumpy/no patches; 
\item[0.25:] 1-2 clumps but no patches OR no clumps but some patches; 
\item[0.5:] 3+ clumps but no patches OR 1-2 clumps but some patches OR no clumps but lots of patches; 
\item[0.75:] 3+ clumps and some patches OR 1-2 clumps and lots of patches; 
\item[1.0:] 3+ clumps and lots of patches.  
\end{enumerate} For each galaxy, each inspector
can choose more than one option (cell) based on whether the galaxy has any blue
patches (diffuse discrete regions based on the F606W image) and/or any major
clumps (concentrated discrete regions based on both F606W and F160W images).
The scores of chosen cells are then averaged to a single score of the
inspector. The single scores from all inspectors for the galaxy are averaged to
get the final score between 0 and 1 of the galaxy. This value, visual
clumpiness and patchiness (VC+P), includes both clumps and patches under the
assumption that both are the same phenomenon and have the same physical nature
\citep{trump14}. This assumption is, however, under debate,
as some astronomers believe that the clumps and the patches have different
formation mechanisms. To provide a comprehensive comparison, we measure two
other values from the visual classification. The first one is the score
averaged over only the cells concerned in the clumpiness (Visual Clumpiness or
VC), and the second one is that averaged over only the cells concerned in the
patches (Visual Patchiness or VP).


We compare the number of blobs (NB) of the GOODS-S galaxies in our sample with
VC (left panel), VP (middle), and VC+P (right) in Figure \ref{fig:vvsc}.
There is good agreement (both median and mean) in all three panels,
demonstrating that, in terms of detecting irregular star formation patterns,
our blob finder works consistently with the CANDELS visual inspection.
Interestingly, the scatter in NB vs. VC+P and NB vs. VP is smaller than that in
NB vs. VC, suggesting that our blob finder has better agreement with the visual
values that contain patches. This is not too surprising, though, because our
detection algorithm has no constraints on the concentration of blobs, while VC
does.  The implementation of such a concentration requirement is actually hard
and uncertain in both automated and visual classifications, especially at high
redshift, where the concentration measurement is difficult even for overall
galaxies, let alone for the galactic sub-structures.

If we choose the visual values (VC, VP, or VC+P) equal to 0.25 or the number of
blobs (NB) equal to 1.5 (yellow lines in the figure) as the threshold of being
a ``clumpy'' (more accurately, blobby) galaxy, the agreed classification rate
is $\sim$75\% for all VC, VP, and VC+P. 

If we assume that our automated detection is correct, the fraction of the Type
I error (NB$<$1.5 and visual values$>$0.25, namely, the visual value falsely
accepts a non-clumpy galaxy as a clumpy galaxy) and the Type II error (NB$>$1.5
and visual values$<$0.25, namely, the visual value falsely excludes a clumpy
galaxy as a non-clumpy galaxy) show opposite behaviors for VC and VP. For VC,
the Type I error (10\%) is less than the Type II error (15\%), while for VP,
the Type I error (20\%) is larger than the Type II error (5\%).  
This interesting result gives us a guideline of using the VC and VP: if one
wants a conservative sample only focusing on well defined clumps, VC should be
used to reduce the Type I error, although it may miss some diffused clumps. On
the other hand, if one wants a sample with as many as possible clumpy
candidates, VP or VC+P should be used to reduce the Type II error.

\end{document}